\documentclass[11pt,a4paper]{article}
\usepackage[a4paper, total={6in, 8in}]{geometry}
\usepackage{standalone}
\usepackage{blindtext}
\usepackage{graphicx} 
\usepackage{physics}
\usepackage{graphicx}
\usepackage{dcolumn}
\usepackage{bm}
\usepackage{xcolor}
\usepackage[utf8]{inputenc}
\usepackage[T1]{fontenc}
\usepackage{mathptmx}
\usepackage{etoolbox}
\usepackage{csquotes}
\usepackage[british]{babel}
\usepackage{amsmath}
\usepackage{siunitx}
\usepackage[]{mdframed}
\renewcommand{\thepage}{\arabic{page}}
\usepackage{caption}
\usepackage{romannum}

\begin{document}
\title{Probing of magnetic dimensional  crossover in CrSiTe$_{3}$ through picosecond strain pulses}
\date{} 
\maketitle
Anjan Kumar N M$^{1}$, Soumya Mukherjee$^{1}$, Abhirup Mukherjee$^{1}$,
Ajinkya Punjal$^{2}$, Shubham Purwar$^{3}$, Thirupathaiah Setti$^{3}$,
Shriganesh Prabhu S$^{2}$, Siddhartha Lal$^{1}$, N. Kamaraju$^{1*}$ \\

Corresponding authors email : nkamaraju@iiserkol.ac.in\\

$^{1}$ Department of Physical Sciences, Indian Institute of Science
Education and research, Kolkata, 741246, West Bengal, India.\\
$^{2}$ Department of Condensed Matter Physics and Materials Science, Tata
Institute of Fundamental research, Mumbai, 400005, Maharashtra, India.\\
$^{3}$ Department of Condensed Matter and Materials Physics, S. N. Bose
National Centre for Basic Sciences, Kolkata, 700 106, West Bengal, India

\section*{Abstract}
Elucidating the emergence of long-range magnetic ordering from its precursor short-range magnetic ordering (SRMO) in two-dimensional van der Waals materials holds profound implications for fundamental research and technological advancements. However, directly observing the intricate stages of this magnetic dimensional crossover (MDC) remains a significant experimental challenge. While magneto-elastic coupling offers a promising avenue, detecting the minute lattice response to SRMO proves challenging. Recent investigations utilizing second harmonic generation have unveiled a two-step MDC in a van der Waals ferromagnetic insulator. However, an unambiguous detection of MDC through the  time-resolved techniques remains elusive. To meet this goal, we have executed an alternative approach by employing picosecond acoustic strain pulses generated by femtosecond lasers to probe the various stages of MDC through the magneto-elastic coupling for the first time. By analyzing the shape  of the strain pulse in both the time and frequency domains as a function of temperature, we clearly demonstrate  the detection of  the subtle influence of spin fluctuations on the lattice. Additionally, the ultrafast carrier dynamics also show signatures of MDC. Our measurements pave the way towards characterizing magnetic materials in time-resolved experiments that are crucial in designing a new generation of spin-based optoelectronic devices.

\section*{Introduction}
Unveiling the impact of long-range magnetic ordering (LRMO) on lattice, charge, and orbital degrees of freedom in magnetic materials is crucial from both basic and technological standpoints\cite{burch2018magnetism, li2019intrinsic}.  There has been a surge of interest in understanding the multistep process by which spin fluctuations drive the magnetic dimensional crossover in van der Waals materials\cite{huang2017layer, gong2017discovery, girovsky2017long} from two-dimensional (2D) short-range magnetic ordering to three-dimensional LRMO. Such a MDC is necessitated by the Mermin–Wagner–Hohenberg theorem\cite{mermin1966absence,PhysRev.158.383}. Recently, a two-step MDC has been detected in a van der Waals ferromagnetic (FM) Insulator with strong magneto-elastic coupling, CrSiTe$_{3}$, using second harmonic generation (SHG) experiments\cite{ron2019dimensional} that are sensitive to minute atomic distortions. The MDC is observed to proceed from a high-temperature paramagnetic phase with 2D SRMO fluctuations to a low-temperature three-dimensional (3D) ferromagnetically ordered phase, through an intervening phase that displays 2D magnetic LRMO fluctuations. Since the atomic distortions caused by SRMO are difficult to detect in X-ray, neutron, and optical scattering, studies have typically been carried out using Raman active modes \cite{casto2015strong, milosavljevic2018evidence}. A recent investigation using ultrafast time-resolved measurements\cite{ron2020ultrafast} has examined the influence of SRMO fluctuations on coherent A$^3_{g}$ optical phonon. However, achieving a clear detection of the entire voyage of magnetic dimensional crossover (MDC) through its various stages using time-resolved techniques continues to be a challenge.

Here, we study the influence of strong magnetoelastic coupling in CrSiTe$_{3}$ by employing ultrafast femtosecond pulses to generate a stress composed of thermoelastic \cite{thomsen1986surface} , electronic \cite{wu2007femtosecond} and magnetic origin \cite{von2020unconventional, mattern2023towards} in addition to producing photoexcited carriers\cite{lovinger2020magnetoelastic, fu2022strain}. In turn, this stress generates a picosecond acoustic strain pulse that will be sensitive to MDC in magnetic materials. Tracking the picosecond acoustic strain pulse in both the frequency and time domains, as a function of temperature, we detect the various stages of MDC. Additionally, we observe that the ultrafast dynamics of the photoexcited carriers too show signatures of MDC. Consequently, the findings of this study offer a valuable tool for examining the influence of magnetic ordering on the dynamics of photoexcited carriers, a critical aspect in the development of novel spin-based optoelectronic devices. To the best of our knowledge, this is the first time such a method has been employed to investigate MDC in magnetic materials.

In this work, the ultrafast dynamics of photoexcited electrons and the generated strain pulse characteristics are  investigated in bulk CrSiTe$_{3}$ at various temperatures using a non-degenerate pump-probe technique(see Fig. 1(a) for the schematic). The transient differential reflectivity data reveal a pump-induced acoustic strain pulse superimposed on an exponentially decaying electronic background. The relaxation timescales and coefficients extracted from multiexponential data fitting to the electronic background decay exhibit a significant change around the temperature T$_{3D}$  $\sim$ 50 K, where the onset of LRMO fluctuations begins \cite{ron2019dimensional}. In addition, the shape of the acoustic strain pulses exhibits a continuous change across each stage of magnetic dimensional crossover.
By analyzing the strain pulses in the frequency domain, it becomes evident that the onset of LRMO fluctuations, brought about by the emergence of interlayer spin-spin interactions, impacts the acoustic strain pulses. This impact manifests itself as a softening of the high-frequency acoustic part by  $\sim$ 6 GHz and a gapping out of the low-frequency acoustic part by $\sim$ 7 GHz. Both these behaviors and the change in the phase of the strain pulse are effectively captured by our phenomenological model, which takes into account the coupling of magnetic interactions, with lattice distortion and its gradient.



\section*{\textbf{Experiments $\&$ Results}}
\begin{figure*}%
\includegraphics[width= 1\textwidth]{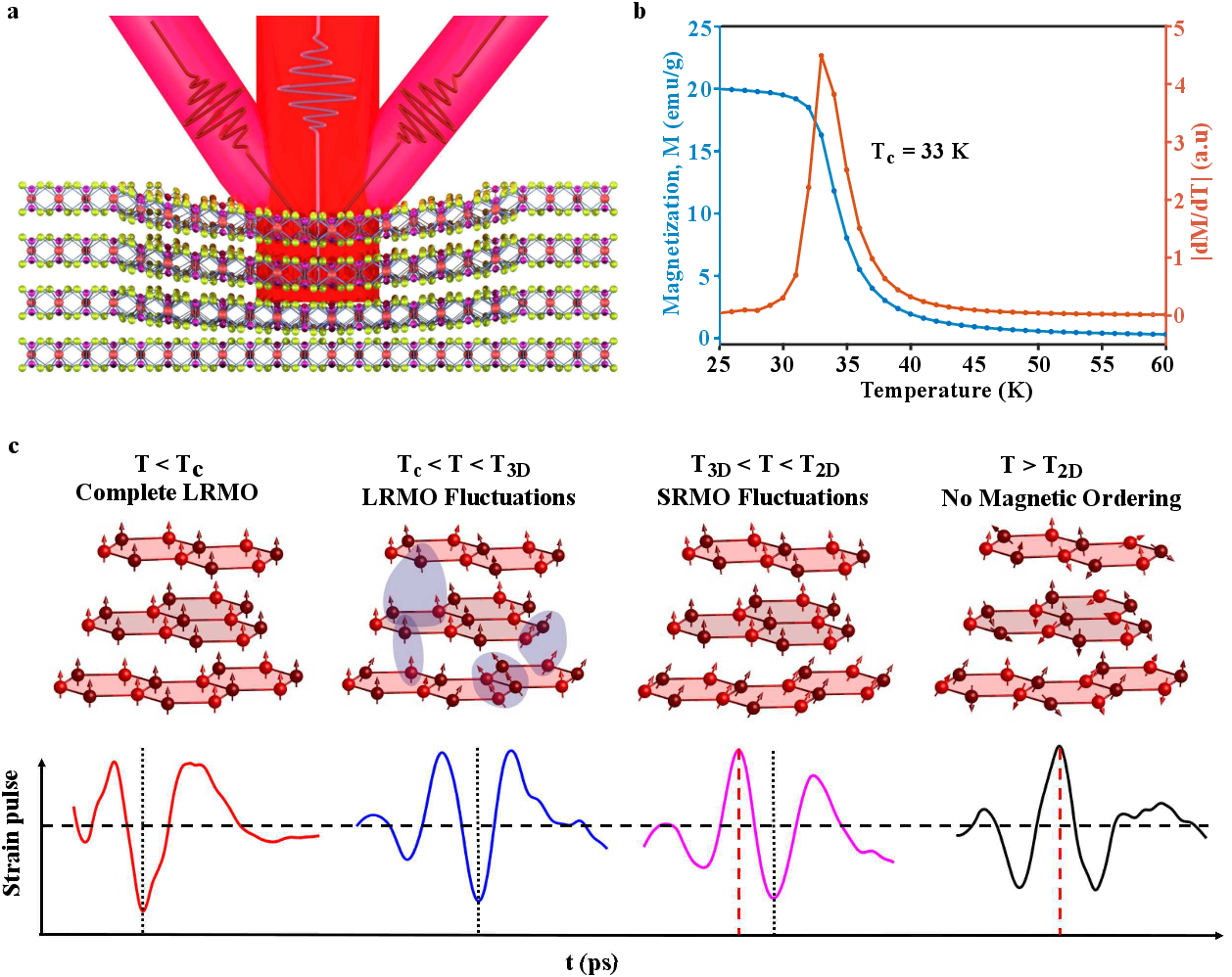}
\caption{ Detection of dimensional crossover using strain pulses: (a) Schematic of ultrafast pump probe technique to generate and detect the strain pulse (b) Variation of magnetization as a function of temperature (blue filled hexagons are the data and the blue lines are the guide to the eye) and absolute value of derivative of  magnetization (dM/dT) versus temperature (orange lines are the guide to the eye, connecting the orange filled hexagons(data)). Transition temperature is found here to be 33 K, (c) The various stages of magnetic ordering captured by the strain pulse, depicted along with the schematic of $\lq$Cr' spin ordering in the layers. One can notice that the maximum of the strain pulse is negative at T $<$ T$_{c}$ and the shape of the strain pulses gradually changes to a stage where the maximum of the pulse becomes positive at T $>$ T$_{2D}$. The horizontal black dashed line denotes the reference level (y = 0). Vertical lines, distinguished by dotted and dashed styles, identify the extrema of the strain pulses. The red dashed line marks the tensile strain (positive peak), while the black dotted line marks the compressive strain (negative peak). Note that at T $>$ T$_{2D}$ 
, the maxima and minima are comparable and hence both extremas are marked. }\label{all_processes}
\end{figure*}

CrSiTe$_{3}$ is a layered ferromagnet with Cr atoms forming honeycomb lattices. The ferromagnetic exchange interactions between the Cr ions result in spontaneous LRMO below the Curie temperature of 33 K (refer to Fig. \ref{all_processes} (b)), with the easy axis aligned along the c-axis \cite{casto2015strong, williams2015magnetic}.The commencement of the LRMO fluctuations through the interlayer spin-spin interaction  is known to begin at  $\sim$ 60 K based on SHG investigations\cite{ron2019dimensional}. Inelastic neutron scattering experiments have revealed that CrSiTe$_{3}$ has intralayer spin-spin correlations even up to 300K, which are prominent below $\sim$ 110 K\cite{williams2015magnetic}. As a result, the temperature at which the in-plane spin-spin interaction appears is denoted as T$_{2D}$. The multi-step energetics to establish the LRMO in CrSiTe$_{3}$ \cite{ron2019dimensional}, and the presence of strong spin-lattice coupling\cite{casto2015strong,milosavljevic2018evidence},  makes this material an ideal test bed to investigate the spin-spin correlations and its effect on the lattice using picosecond acoustic strain pulses(see Fig. \ref{all_processes} (c)). The various processes listed in Fig. 1(c) will be discussed further in the experimental results.

\begin{figure*}%
\centering
\includegraphics[width=1\textwidth]{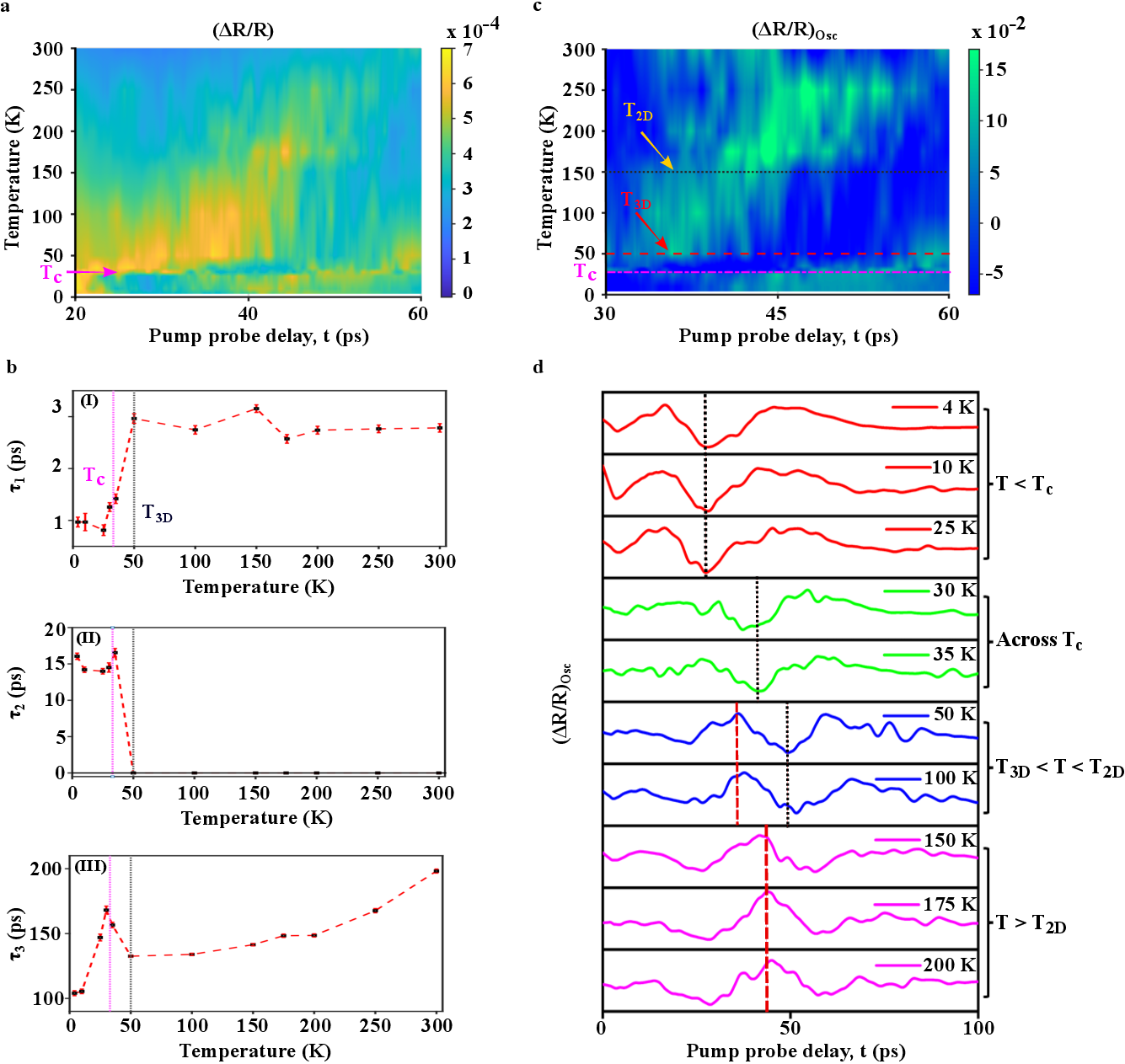}
\caption{Effect of magnetic order on the carrier dynamics and the strain pulses: (a) A 2D false color plot of temperature dependent evolution of raw transient reflectivity data. Curie temperature (T$_{c}$) is indicated in solid red arrow. For clarity, the data is shown up to the pump probe delay time from 20 ps to 60 ps. (b) relaxation timescales extracted from a multi-exponential fit to the transient differential reflectivity data are shown as black squares, with error bars determined using orthogonal distance regression (ODR) approach. The red dotted lines are guides to the eye. The vertical magenta and black solid lines indicate the Curie temperature (T$_{c}$) and the temperature (T$_{3D}$) at which the onset of LRMO  begins, respectively. (c) A 2D false color map of strain pulses is shown as a function of pump-probe delay and temperature. The horizontal black dotted line, red dashed line, and magenta dot-dashed line mark the temperatures T$_{2D}$, T$_{3D}$, and T$_{c}$, respectively. For clarity, the data is shown up to the  pump probe delay time of 30 ps to 60 ps (d) The extracted strain pulses from DR data  at various temperatures are shown in different colors. Red, green, blue and magenta solid lines represent strain pulses at temperatures T $<$ T$_{c}$, across T$_{c}$, T$_{3D}$ $<$ T $<$ T$_{2D}$ and T $>$ T$_{2D}$, respectively. Vertical lines, distinguished by dotted and dashed styles, identify the extrema of the strain pulses. The red dashed lines mark the maximum tensile strain (positive peak), while the black dotted lines mark the minimum compressive strain (negative peak)}\label{DR_data_parameters}
\end{figure*}

In this ultrafast pump probe experiment, an ultrashort visible pulse ($\sim$ 60 fs, $\sim$ 1.9 eV) excites the sample by transferring the electrons directly from Te 5p level to Cr 3d level (see Supplementary Sect. II and Fig. S6). This charge transfer significantly enhances the ferromagnetic super-exchange interaction between the Cr$^{3+}$ metal ions, which are mediated by a nonmagnetic Te ligand\cite{ron2020ultrafast}. To gauge pump-induced changes, a probe pulse with a photon energy centered at $\sim$ 1.55 eV is used to measure the transient differential reflectivity (DR) at various temperatures ranging from 4 K to 300 K. The temperature evolution of the raw DR data is shown in Fig. \ref{DR_data_parameters}(a) as a two dimensional (2D) false colour plot (refer to Supplementary Fig. S5(a-c) for the complete data). The raw DR data shows an abrupt change near the Curie temperature (T$_{c}$ $\sim$ 33K, refer Fig. \ref{DR_data_parameters}(a)).The photoinduced DR data is positive after the pump pulse excites the electrons to the excited state (refer to Supplementary Fig. S 5(b-c)). These photoexcited electrons exchange energy with the spin and lattice subsystems, each with a characteristic relaxation timescale reflected in the exponential decay of the DR data background. On top of the exponentially decaying background the raw DR data has a strain pulse (refer to Supplementary Fig. S 5(b-c)). The decaying exponential part of the DR data could be successfully fitted with the multi-exponential functions. It should be noted that a minimum sum of three exponentials is required to fit the data taken up to the temperature of 35K, whereas, only two exponentials are sufficient to fit the  exponential background of  DR data taken at the temperatures T $>$ 35K (refer to Supplementary Fig. S7(a-b)). The fact that the electronic background of DR is found to be fitted with a different number of exponentials across 35 K suggests that LRMO or the onset of LRMO fluctuations influence the electron dynamics in CrSiTe$_{3}$.  To further assert our claim, we examine the effect of the spin fluctuations on the decay time constants and the decay coefficients of the electronic background. The decay time constants extracted from the multi-exponential fitting to the electronic background and its variations with the temperature is shown in Fig. \ref{DR_data_parameters}(b). The associated decay coefficients are shown in Supplementary Fig. S9.

The values of the extracted  decay time constants($\tau_{1}\sim 1-3$ ps , $\tau_{2}\sim 15$ ps, $\tau_{3}\sim 100-200$ ps), are well-separated and can be attributed to different physical processes as discussed below. The pump pulse starts the Te - Cr charge transfer process by creating hot electrons in the conduction band. These hot electrons then thermalize through electron-electron interactions and then exchange energy with the lattice and spin degrees of freedom. The fastest relaxation timescale, $\tau_{1}$, which is shown in Fig. \ref{DR_data_parameters}(b), increases from $\sim$ 1.5 ps to $\sim$ 3 ps as the temperature rises from 4 K to 50 K, and it is nearly independent of temperature above 50 K to 300 K. This decay is attributed to electron-phonon thermalization \cite{guo2020anomalous}. As $\tau_{1}$ is $\sim$ 1.5 ps below 50 K and 3 ps above 50 K, it indicates that an extra scattering channel is activated below T$_{3D}$ $\sim$ 50 K, and this could be the spin channel \cite{guo2020anomalous, liu2022electron, sutcliffe2023transient}. 

Subsequent to the energy exchange between the electron and phonon subsystems, the lattice-spin coupling induces an additional energy and angular momentum exchange between the lattice and spins \cite{beaurepaire1996ultrafast, lovinger2020magnetoelastic, lichtenberg2022anisotropic}. This process is characterized by the relaxation timescales $\tau_{2}$ and $\tau_{3}$ . The timescale $\tau_{2}$, emerging only below 50 K (T$_{3D}$), suggests a link to the  interaction of phonons with interlayer spin-spin fluctuations. The value of approximately 16 ps for $\tau_{2}$ aligns with the typical time scales of spin-lattice interaction observed in another 2D ferromagnetic van der Waals material, Fe$_{3}$GeTe$_{2}$\cite{lichtenberg2022anisotropic, li2022abnormal}. On the contrary, the time scale $\tau_{3}$ ranges from 100 to 200 ps and shows a peak at the Curie temperature (T$_{c}$). This third timescale is attributed to the lattice interaction with the intralayer spin-spin fluctuations, in keeping with the observation of a divergence in the intralayer spin-spin correlation length at T$_{c}$ in neutron diffraction studies\cite{williams2015magnetic}. Furthermore, $\tau_{3}$ increases as the sample is heated up: this is likely due to the decrease in the strength of intralayer spin-spin correlations and a concomitant reduction in its interaction with the lattice. Thus, CrSiTe$_{3}$ exhibits two distinct lattice-spin coupling mechanisms: one linked to the interlayer spin-lattice interaction (which is dominant below 50 K) and another originating from the intralayer spin-lattice interaction. Both highlight the complex interplay between lattice and spin dynamics in this  material.
Alongside the triexponential function, there exists a constant background (A$_{0}$) and this can be attributed to a long electron-hole recombination time. A$_{0}$ also shows a peak at T$_c$ (see Supplementary Fig. S9 (d)), suggesting that the electron-hole recombination process is affected by magnetic ordering\cite{yan2023extending}.

Next, we turn our attention to the intriguing oscillatory part of the DR data. We retrieve the strain pulse (oscillatory component) by subtracting the multi-exponential electronic background at each temperature.  The 2D contour plot of time dependent strain pulses is shown in Fig. \ref{DR_data_parameters} (c).  The essential aspect to consider here is the effect of temperature on the shape of the strain pulse. Figure \ref{DR_data_parameters}(d) depicts evolution of the time domain strain pulses (smoothened data for clarity, see Supplementary Sec. $\Romannum{4}$ and Fig. S11) at different temperatures upto 200 K. The raw data at all temperatures is shown in Supplementary Fig. S10. From Fig. \ref{DR_data_parameters}(d) it is evident that every stage of the spin-spin correlations (dimensional crossover) is affecting the shape of the strain pulse. In particular, the strain pulse in the ferromagnetic (FM) phase (red solid lines in Fig. \ref{DR_data_parameters} (d)) is inverted compared to the strain pulse in the paramagnetic (PM) phase (magenta solid lines in Fig. \ref{DR_data_parameters} (d)). For the clarity the data at FM (4K) and PM (150 K) phases is shown in the Supplementary Fig. S12. 

\begin{figure*}
    
\centering
\includegraphics[width=1\textwidth]{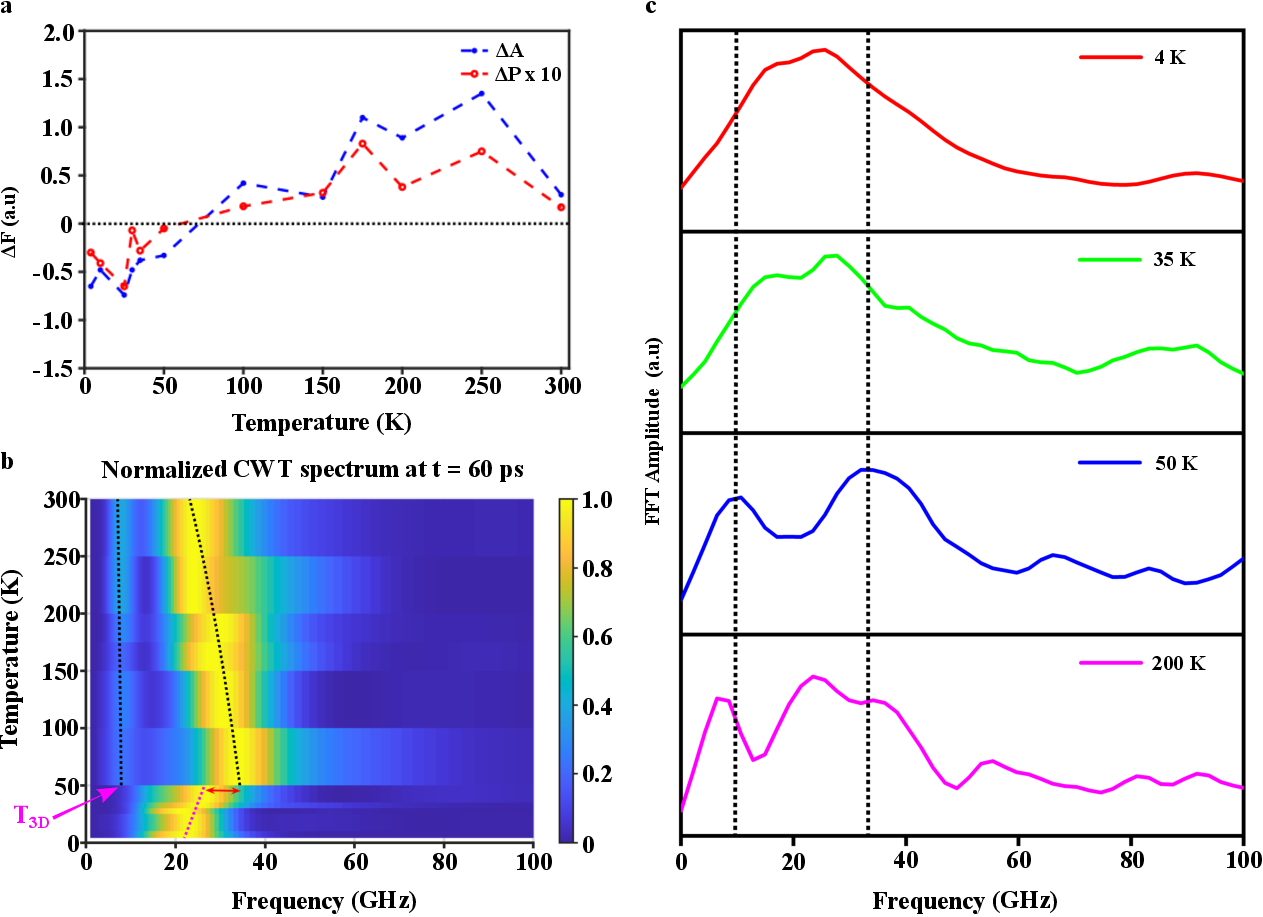}
\caption{(a) The variation of the difference in the peak-to-peak value ($\Delta$P) of the strain pulses and the difference in the area under the curve of the positive and negative sides of the strain pulse ($\Delta$A) as a function of temperature. For the clarity, both  ($\Delta$P) and ($\Delta$A) are represented as $\Delta$F(T), and ($\Delta$P) is scaled up by a factor of 10. (b) The 2D false color map of evolution of the frequency spectrum obtained by continuous wavelet transform of the strain pulse with the temperature at the pump probe delay time of 60 ps. The data is normalized with respect to its peak value and shown up to 100 GHz for clarity.  The variation in frequency of LFAP and HFAP with the temperature above the onset of LRMO temperature T$_{3D}$ $\sim$ 50 K is fitted with the anharmonic decay model and is shown here as the dotted black lines. The dotted red lines are guide to the eye, to indicate the softening of HFAP below T$_{3D}$.  A solid red double arrow depicts the shift in the central frequency of HFAP across T$_{3D}$ . (c) This plot shows Fourier-transformed frequency spectra at four  temperatures ( only 4 data are shown here for clarity). Data are smoothed for better visualization without affecting peak frequencies. Vertical dotted lines mark the central frequencies of LFAP and HFAP, shown relative to the 50 K spectrum. Noticeably, above T$_{3D}$ = 50 K, both peaks redshift (move towards lower frequencies). However, below T$_{3D}$, HFAP red-shifts while LFAP blue-shifts (moves towards higher frequencies). Our main text explains this contrasting behavior through spin-lattice coupling.  }\label{Area_frequency}
\end{figure*}

The shape of strain pulses produced by femtosecond pulses in semiconductors is influenced by the interplay between thermoelastic (TE) and deformation potential (DP) induced stresses \cite{thomsen1986surface, wu2007femtosecond, von2020unconventional, mattern2023towards}. These stresses produce a bipolar strain pulse with a leading compressive and a trailing tensile part. When a femtosecond pulse is shone on a magnetic semiconductor such as CrSiTe$_{3}$, magnetic stresses are also produced in addition to TE and DP induced stresses. This magnetic stress generate a contractive strain pulse \cite{ von2020unconventional, mattern2023towards}. The inversion of the strain pulse in the ferromagnetic phase compared to the paramagnetic phase suggests an interplay between these different generation processes: i.e, in  FM phase, the strain pulse has a more negative amplitude than the positive amplitude, while in the PM phase, the positive  amplitude dominates the negative amplitude (refer to Supplementary Fig. S12). We provide an explanation of the inversion of the strain pulse below based on a phenomenological model of spin-lattice coupling. To quantify the interplay between these processes, the difference in the peak-to-peak value ($\Delta$P) of the strain pulses and the area under central part of the strain pulse ($\Delta$A) are extracted (The extraction process is depicted in the Supplementary Fig. S13). Both $\Delta$P and $\Delta$A, show a similar trend with temperature in Fig. \ref{Area_frequency} (a). In Fig. \ref{Area_frequency} (a), $\Delta$F(T) represents both $\Delta$P and $\Delta$A as a function of temperature (T). $\Delta$F(T) decreases with decreasing temperature, especially below 150K, and changes sign around a temperature of T$_{3D}$ $\sim$ 50 K. This behavior can be explained as follows: In the paramagnetic phase, the positive  $\Delta$F(T) indicates a dominance of positive strain contributions, indicating the lattice expansion. This expansion results from elastic stresses induced by the TE and DP processes. Upon cooling, below 150 K, the strengthening spin-spin correlations generate a contractive strain pulse through the magnetostriction effect. This counteracts the elastic expansive stress, diminishing the positive portion of $\Delta$F(T). Furthermore, below 50 K, the emergence of long-range magnetic order (LRMO) fluctuations leads to a non-zero net magnetization (M). This non-zero magnetization further enhances the contractive strain pulse, causing  $\Delta$F(T) to exhibit a negative sign. Supporting these observations, studies on magnetic thin films with optical pump-X-ray probe methods have hinted at the interplay between elastic and magnetic stresses\cite{schmising2008ultrafast,von2020unconventional,von2020spin}.






To understand these observations further, we resort to Fourier and wavelet analysis of the time domain strain data. The continuous wavelet transform (CWT) effectively captures the evolution of the temperature-dependent frequency spectrum at various pump-probe time delays (Refer to Supplementary section $\Romannum{5}$).   Figure \ref{Area_frequency}(b) presents the temperature evolution of the frequency spectrum (normalized with respect to the peak amplitude obtained from wavelet analysis) at the pump probe delay time, t = 60 ps, as a 2D false color plot (refer to the Supplementary Figs. S15 and S16 for the evolution of the CWT spectrum at various pump probe delay times and at various temperatures). For clarity, smoothened frequency spectra obtained from Fourier transform at four  temperatures are shown in Fig. \ref{Area_frequency}(c). The frequency spectrum has two parts: a high-frequency acoustic part (HFAP) and a low-frequency acoustic part (LFAP) . The HFAP's central frequency undergoes a significant renormalization, dropping from 33 GHz to 27 GHz (marked by the double red arrow in  Fig. \ref{Area_frequency}(b)). Below T$_{3D}$, it softens markedly and exhibits a red shift above T$_{3D}$. On the contrary, the LFAP emerges above T$_{3D}$ and shows a weak temperature dependence. Interestingly, below T$_{3D}$, the LFAP undergoes a blue shift by 7 GHz (toward higher frequencies), which leads to its disappearance (gapping out) within this temperature regime.


The influence of spin-spin correlations on the shape and the frequency of the strain pulse observed by us can be elucidated by employing a phenomenological model Lagrangian given by,

\begin{equation}
\mathcal{L} = \sum_{j} \left[ \dot{Q_{j}}^{2} - \lambda(Q_{j+1} - Q_{j})^{2} + \gamma \vec{S}_{j} \cdot \vec{S}_{j+1} Q_{j}  - \xi \vec{S}_{j} \cdot \vec{S}_{j+1} Q_{j}^{2} + \chi \vec{S}_{j} \cdot \vec{S}_{j+1}(Q_{j+1} - Q_{j})^{2}  \right] \label{eq 1}
\end{equation}

The first two terms of the Lagrangian represents the lattice model of phonons, where $Q_{j}$ is the distortion of the $j^{th}$ atom away from its equilibrium position, and $\lambda$ is the spring constant of the restoring potential. The distortions $Q_{j}(t)$ are functions of the position index j as well as time t. The Lagrangian includes additional terms that describe the interaction between lattice distortions and magnetic interactions. These magnetic interactions are assumed to be of the nearest-neighbor kind $\vec{S}_{j} \cdot \vec{S}_{j+1}$, and they can couple to the linear, quadratic, and gradient components of the distortion, represented by $Q_{j}$ , $Q_{j}^{2}$ and $(Q_{j+1} -  Q_{j})^{2}$, through the coupling constants $\gamma$,  $\xi$ and $\chi$, respectively.

In this approach, the magnetic interactions are treated at the mean-field level and are replaced by their expectation values. This leads to the correlation \textlangle $\vec{S}_{j} \cdot \vec{S}_{j+1}$\textrangle  acquiring a uniform non-zero value M$^{2}$(T), where M is the magnetization. When the temperature exceeds T$_{3D}$, M$^{2}$(T) becomes zero. The Lagrangian for the LRMO phase takes the form

\begin{equation}
\mathcal{L} = \sum_{j} \left[ \dot{Q_{j}}^{2} - (\lambda - M^{2}\chi) (Q_{j+1} - Q_{j})^{2} - M^{2}\xi \left(Q_{j} - \frac{\gamma}{2\xi}\right)^{2} \right] \label{eq 2}
\end{equation} where we have neglected the overall constant.

We define a modified distortion $\bar{Q_{j}}$ = $Q_{j} - \frac{\gamma}{2\xi}$. In terms of the modified distortion the Lagrangian becomes, 

\begin{equation}
\mathcal{L} = \sum_{j} \left[ \bar{\dot{Q_{j}}}^{2} - (\lambda - M^{2}\chi) (\bar{Q}_{j+1} - \bar{Q}_{j})^{2} - M^{2}\xi \bar{Q}_{j}^{2} \right] \label{eq 3}
\end{equation}

The above Lagrangian yields the phonon dispersion relation given by,

\begin{equation}
\omega(k) = \sqrt{(\lambda - M^{2}\chi)k^{2} + M^{2} \xi} \label{eq 4}
\end{equation}

where k is the phonon wave-vector. As a consequence, when M is zero (for T $>$ T$_{3D}$), the dispersion reduces to the linear form expected for acoustic phonons: $\omega(k)$ = $\sqrt{\lambda} k$. When M and the coupling $\chi$ are nonzero (for T $<$  T$_{3D}$) but the coupling $\xi$ is negligible, the dispersion has the same form, but the spring constant $\lambda$ is reduced, leading to softening of the mode. 

\begin{equation}
\omega(k) = \sqrt{\lambda_{eff}} k, ~ \lambda_{eff} = \lambda - M^{2}\chi .
\end{equation}

Therefore, the coupling between the spin-spin correlations and the gradient component of the distortions leads to the softening of the HFAP across T$_{3D}$ for the case M$^2\chi < \lambda$. 

When M, $\chi$ and $\xi$ are all non-zero (for T $<$ T$_{3D}$), the dispersion relation takes  the form of equation (\ref{eq 4}) which indicates that the coupling between the spin-spin correlations and the quadratic  component of the distortions can lead to the gapping out of the acoustic modes below T$_{3D}$ making it harder to observe. This captures the onset of long-range magnetic order (LRMO) fluctuations below T$_{3D}$ observed by us. The observed frequency shift between 50 K and 35 K, with a blue shift of about 7 GHz for LFAP and a red-shift of about 6 GHz for HFAP, provides insights into the parameters $\chi$ and $\xi$. This suggests that both acoustic components could originate from two distinct branches of acoustic phonons ((refer to Supplementary Sec. $\Romannum{7}$). From the analysis (refer to Supplementary Sec. $\Romannum{7}$ and Fig. S17 and S18), the values of  $\chi$  and $\xi$ are found to be $\sim$ 1.7 mHz $m^{2}/A^{2}s$ $\sim$ and 10.2 THz $m^{2}/A^{2}s$, respectively.

Also note that the distortion field $Q$ is modified to $\bar{Q_{j}}$ = $Q_{j} - \frac{\gamma}{2\xi}$ by spin-phonon coupling. After the onset of magnetic order below T$_{3D}$, the amplitude can become negative for large $\frac{\gamma}{2\xi}$, leading to a phase inversion of
the distortions \cite{ergeccen2023coherent}. Above T$_{3D}$, even though there is no finite magnetisation M, $Q$ can still couple with the spin fluctuations $\textlangle (\vec{S}_{j} \cdot \vec{S}_{j+1})^{2}\textrangle$ - $\textlangle \vec{S}_{j} \cdot \vec{S}_{j+1}\textrangle^{2}$, again leading to a reduced amplitude.

Also, above T$_{3D}$  the frequency of both  HFAP as well as the LFAP (although weakly) exhibits a decreasing trend with increasing temperature as expected from anharmonic phonon decay \cite{orbach1964attenuation, klemens1967decay, baumgartner1981spontaneous, tamura1985spontaneous}. The temperature-dependent frequency variation is modeled using the anharmonic phonon decay model and depicted as black dashed lines in Fig. \ref{Area_frequency} (details of the model are available in Supplementary Sec. $\Romannum{5}$). Importantly, the frequency variation of the LFAP with increasing temperature is less pronounced compared to the HFAP, indicating weaker anharmonic coupling in the LFAP. Further discussion on the frequency shift and lifetime of both LFAP and HFAP can be found in Supplementary Sec. $\Romannum{5}$.

\section*{Conclusions}
In conclusion, our comprehensive non-degenerate pump-probe experiments combined with a phenomenological model provide a profound understanding of the intricate interplay between spin, charge and lattice degrees of freedom in the 2D Heisenberg ferromagnet CrSiTe$_{3}$. We unveil the distinct stages of magnetic dimensional crossover and their impact on both ultrafast carrier dynamics and acoustic phonons. Notably, our findings confirm the crucial role of interlayer spin-spin interactions in stabilizing long-range magnetic order at high temperatures. Furthermore, we shed light on the specific influence of different magnetic phases on the acoustic strain spectrum. Remarkably, the long-range magnetic order fluctuations renormalize the frequency of high-frequency acoustic strain part, while also gapping out the low-frequency part.  These intricate observations have provided valuable insights into the microscopic coupling between carriers, spin and lattice degrees of freedom, that are vital in the design of spin-based optoelectronic devices based on 2D ferromagnetic materials like CrSiTe$_{3}$. 

\section*{Methods}
Sample growth and characterization: CrSiTe$_{3}$ single crystals were synthesized via a self-flux technique. High-purity starting materials, Cr (99.999$\%$ from Alfa Aesar), Si (99.999$\%$ from Alfa Aesar), and Te (99.9999$\%$ from Alfa Aesar), were combined in a 1:2:6 molar ratio and placed within a 10 ml alumina crucible. A second crucible with quartz wool was positioned on top of the growth crucible, and the setup was enclosed in a quartz tube filled with high-purity argon gas. The sealed ampoule underwent gradual heating to 1100 \textdegree C over 7 hours, followed by a 15-hour hold at 1100 \textdegree C, and finally cooling to 700 \textdegree C at a rate of 3.2 \textdegree C/h. Upon reaching 700 \textdegree C, excess flux was removed from the crystals by centrifuge technique.

X-ray diffraction (XRD) and energy-dispersive X-ray analysis (EDAX) were employed to analyses the structure and chemical composition of CrSiTe$_{3}$ crystal (Refer Fig. S1 and Tab. TS1 of the Supplementary). From XRD it is confirmed that CrSiTe$_{3}$ crystallizes to R3, space group 148. Temperature dependent powder X-ray diffraction measurements were performed on crushed single crystals using a Rigaku X-ray diffractometer (Smart Lab, 9kW) with Cu K alpha radiation (l = 1.540 Angstrom) over a temperature range of 10-300 K (refer to Supplementary Fig. S2). Raman spectroscopy (Horiba jobin Yvon - HR800, 488 nm excitation) was used to characterize the typical Raman modes of CrSiTe$_{3}$ single crystal (refer to Supplementary Fig. S3). Temperature dependence of the magnetization (M) in field-cooled and zero-field-cooled modes under a magnetic field of 1 kOe applied along the c-axis (refer to Supplementary Fig. S4).

Pump Probe Spectroscopy: Temperature dependent time-resolved non-degenerate transient reflection experiments were used to examine the photoexcited carrier dynamics of CrSiTe$_{3}$ single crystal. The pump-probe setup is set in the reflection geometry. The beam from the Amplifier laser (Spectra Physics Solstice Ace, 800 nm, 1 kHz, 50 fs, 5mJ) was split in two parts. One part of the beam is fed to Optical Parametric Amplifier (Light Conversion TOPAS Prime) to get 650 nm output. The other beam of 800 nm is used for probing the sample. Both the beams are time-delay matched using a retro-reflector on the 600 ps delay stage. The pump beam is modulated at 333 Hz using a mechanical chopper. The change in reflectivity with respect to with and without pump is detected using a balanced photo diode. The sample was cooled using a flow type cryostat (Oxford Instruments) with liquid Helium and hence the sample surface remained in the inert atmosphere all throughout the measurements. The measured full width half maximum spot size of  pump and probe were  $\sim$ 600 $\mu$m and  $\sim$ 400 $\mu$m, respectively. The pump and probe fluences were kept at 1.77 mJ/cm$^{2}$ and 133 $\mu$J/cm$^{2}$ at all the temperatures.


\subsection*{Supplementary information}

Additional discussion on sample characterization (Supplementary Section $\Romannum{1}$), data fitting of pump probe dynamics and  extracted parameters and their temperature dependence (Supplementary Sections $\Romannum{2}$ and $\Romannum{3}$), origin of  picosecond strain pulses and their characteristics in time and frequency domain (Supplementary Sections $\Romannum{4}$ and $\Romannum{5}$), details of theoretical model and the extraction of coupling constants (Supplementary Sections $\Romannum{6}$ and $\Romannum{7}$) and details of data fitting (Supplementary Section $\Romannum{8}$).



\subsection*{Acknowledgments}

The authors thank the Ministry of Education (MoE), Government of India for funding and  IISER Kolkata  for the infrastructural support to carry out the research.
ANM, and SM thank IISER Kolkata and DST-INSPIRE respectively  for their research fellowship. The authors thank Poulami Ghosh and Pedisetti Venkatesh for designing the essential schematics. SL thanks the SERB, Govt. of India for funding through MATRICS grant MTR/2021/000141 and Core Research Grant CRG/2021/000852. N. K thanks the SERB, Govt. of India for funding through Core Research Grant CRG/2021/004885.





\subsection*{Authors' contributions}
A.N.M and N.K. conceived the idea, the project and the analysis. A.N.M, S.M. and A.P.  performed the time-resolved non-degenerate pump probe measurements assisted by S.G.P and N.K.   T.S and S.P grew the crystals and measured the temperature dependent magnetization.  A.M and S.L developed the theoretical model.  A.N.M, and N.K. interpreted the results. A. N.M, S.M., A.M., A.P, S.G.P, S.L, and N. K were involved in the writing of the manuscript. All the authors agree to the content of the final version.



\pagebreak

\setcounter{figure}{0}
\renewcommand{\figurename}{FIG}
\renewcommand{\thefigure}{S\arabic{figure}}
\setcounter{table}{0}
\renewcommand{\tablename}{Table}
\renewcommand{\thetable}{TS\arabic{table}}
\renewcommand\refname{Supplementary References}
\renewcommand \thesection{\Roman{section}}
\renewcommand\thesubsection{\Roman{section}.\alph{subsection}}






\setcounter{figure}{0}
\renewcommand{\figurename}{FIG}
\renewcommand{\thefigure}{S\arabic{figure}}
\setcounter{table}{0}
\renewcommand{\tablename}{Table}
\renewcommand{\thetable}{TS\arabic{table}}
\renewcommand\refname{Supplementary References}
\renewcommand \thesection{\Roman{section}}
\renewcommand\thesubsection{\Roman{section}.\alph{subsection}}
\newpage

\section{Sample characterization}\label{sec1}
\subsection{Single crystal XRD, Energy-dispersive X-ray analysis}

\begin{figure}%
\includegraphics[width= 1\textwidth]{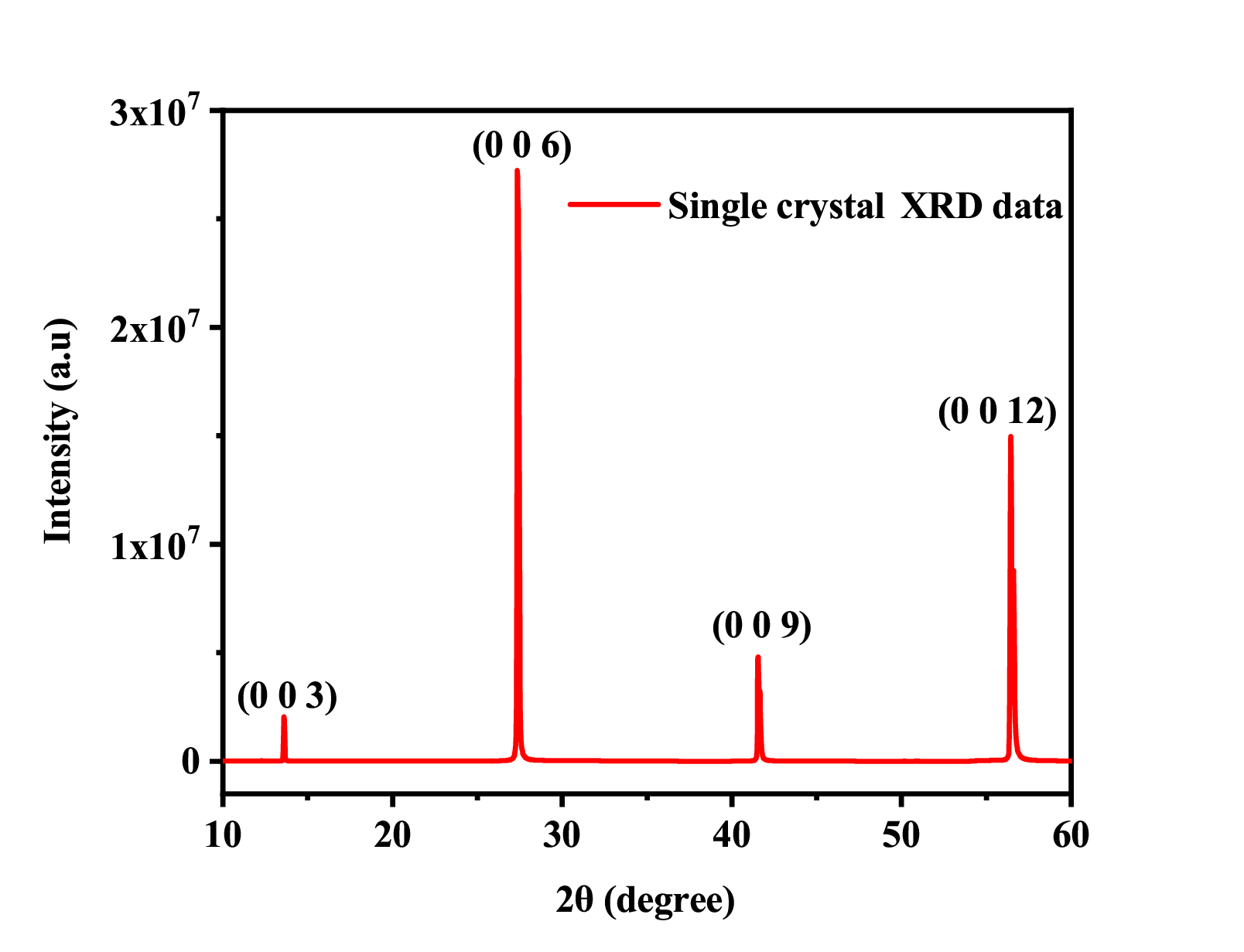}
\caption{ X-ray diffraction spectrum of CrSiTe$_{3}$:  Room temperature X-ray diffraction (XRD) pattern of a single crystal of CrSiTe$_{3}$, showing peaks corresponding to the (003), (006), (009), and (0012) planes. The peaks are consistent with the literature \cite{casto2015strong, milosavljevic2018evidence, zhu2021topological}, indicating the high quality of the crystal. }\label{Sc_XRD}
\end{figure}

 CrSiTe$_{3}$ is a two-dimensional ferromagnetic semiconductor that crystallizes in a rhombohedral structure. The crystal structure consists of three CrSiTe$_{3}$ layers stacked in an ABC sequence, with alternating Te-Cr-Te and Te-Si-Te layers. Within each layer, the Cr atoms form a honeycomb lattice, which is offset from the honeycomb lattices of the adjacent layers \cite{casto2015strong}. The Cr$^{3+}$ ions are situated at the center of a distorted octahedron of six Te atoms (ligand), while the two Si atoms are paired and located between two opposing Te triangles.
Here, X-ray diffraction (XRD) and energy-dispersive X-ray analysis (EDAX) is used to determine the structure and chemical composition of CrSiTe$_{3}$ crystal used in this measurement. Figure \ref{Sc_XRD} shows the XRD data for a single crystal of CrSiTe$_{3}$ at room temperature. The peaks in the XRD data match those reported in the literature \cite{casto2015strong, milosavljevic2018evidence, zhu2021topological}, confirming that the crystal is of high quality and is oriented along the c-axis.

Table \ref{comp} shows the chemical composition of the single crystal CrSiTe$_{3}$.\\

\begin{table}[h]
\caption{Chemical composition of CrSiTe$_{3}$: Exact chemical composition of CrSiTe$_{3}$ single crystal obtained from EDAX is shown here.}\label{comp}%
\begin{tabular}{@{}lll@{}}
\hline\\
Cr (atomic \%) &	Si (atomic \%)&	Te (atomic \%)\\
\hline\\
17.49 & 22.47&	60.03 \\
17.49&	22.17&	60.03 \\
18.06&	22.03	&59.90 \\
19.14&	22.55	&58.41 \\
\hline\\
  Exact composition : & Cr$_{0.9}$Si$_{1.1}$Te$_{2.9}$ & \\
  \hline\\

\end{tabular}
\end{table}

\subsection{Temperature dependent X-ray powder diffraction.}
\begin{figure}%
\includegraphics[width= 0.8\textwidth]{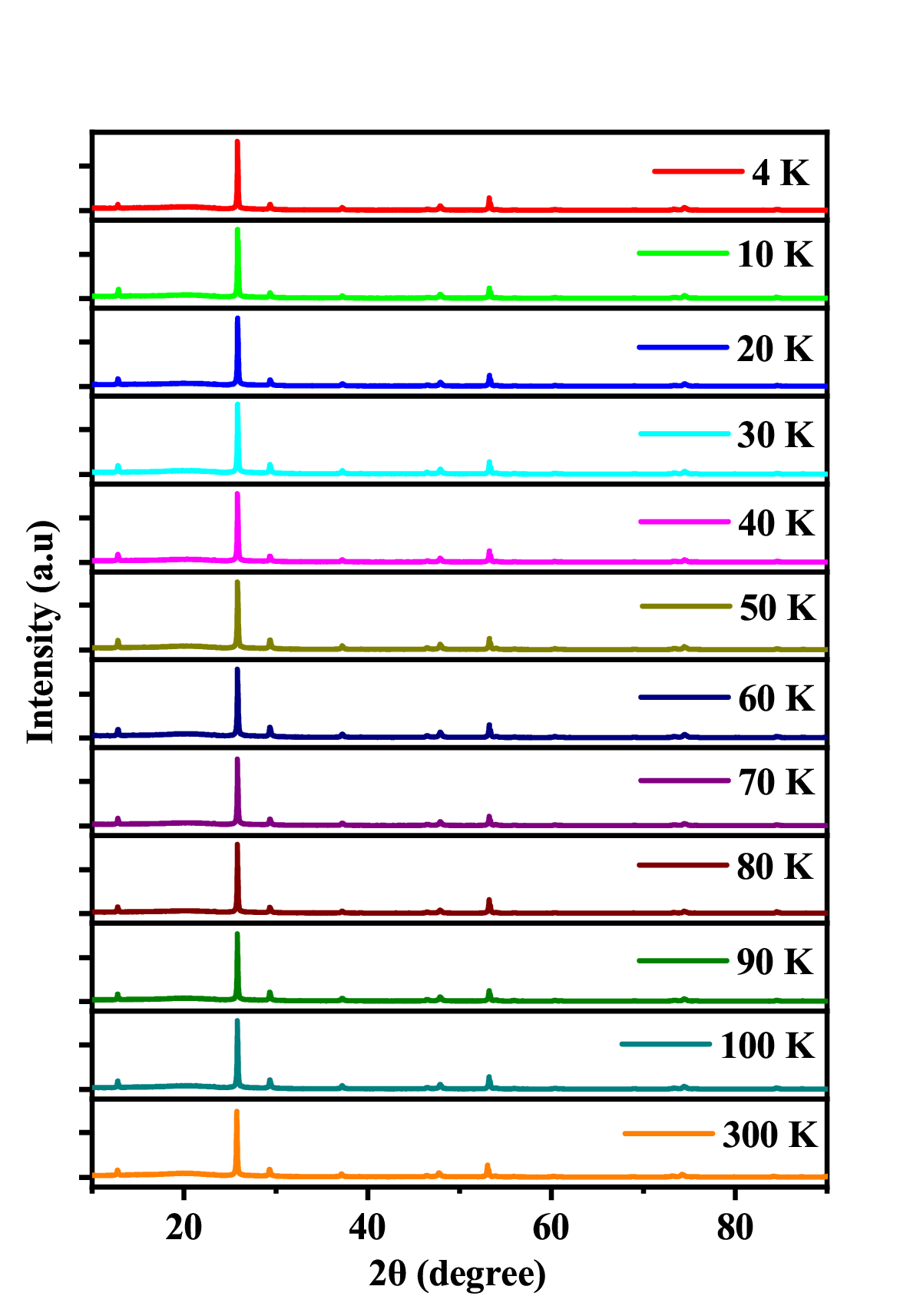}
\caption{Temperature dependent powder x-ray diffraction:  Temperature-dependent powder X-ray diffraction (XRD) spectra of CrSiTe$_{3}$ at various temperatures from 4 K to 300 K. No significant peak shifts or broadening were observed, indicating the absence of a structural phase transition. }\label{T_XRD}
\end{figure}

To investigate any changes in the crystal structure, temperature-dependent powder X-ray diffraction (XRD) measurements were performed on crushed single crystals using a Rigaku Smart Lab 9kW diffractometer with Cu K$\alpha$ radiation (wavelength $\lambda$ = 1.540 $\AA$) over a temperature range of 10-300 K. Supplementary Fig. \ref{T_XRD}. shows the temperature-dependent powder XRD data. No significant peak shifts or broadening were observed across the entire temperature range, confirming the absence of any structural phase transition.

\subsection{Raman Spectroscopy of  CrSiTe$_{3}$ single crystal.}
\begin{figure}%
\includegraphics[width= 1\textwidth]{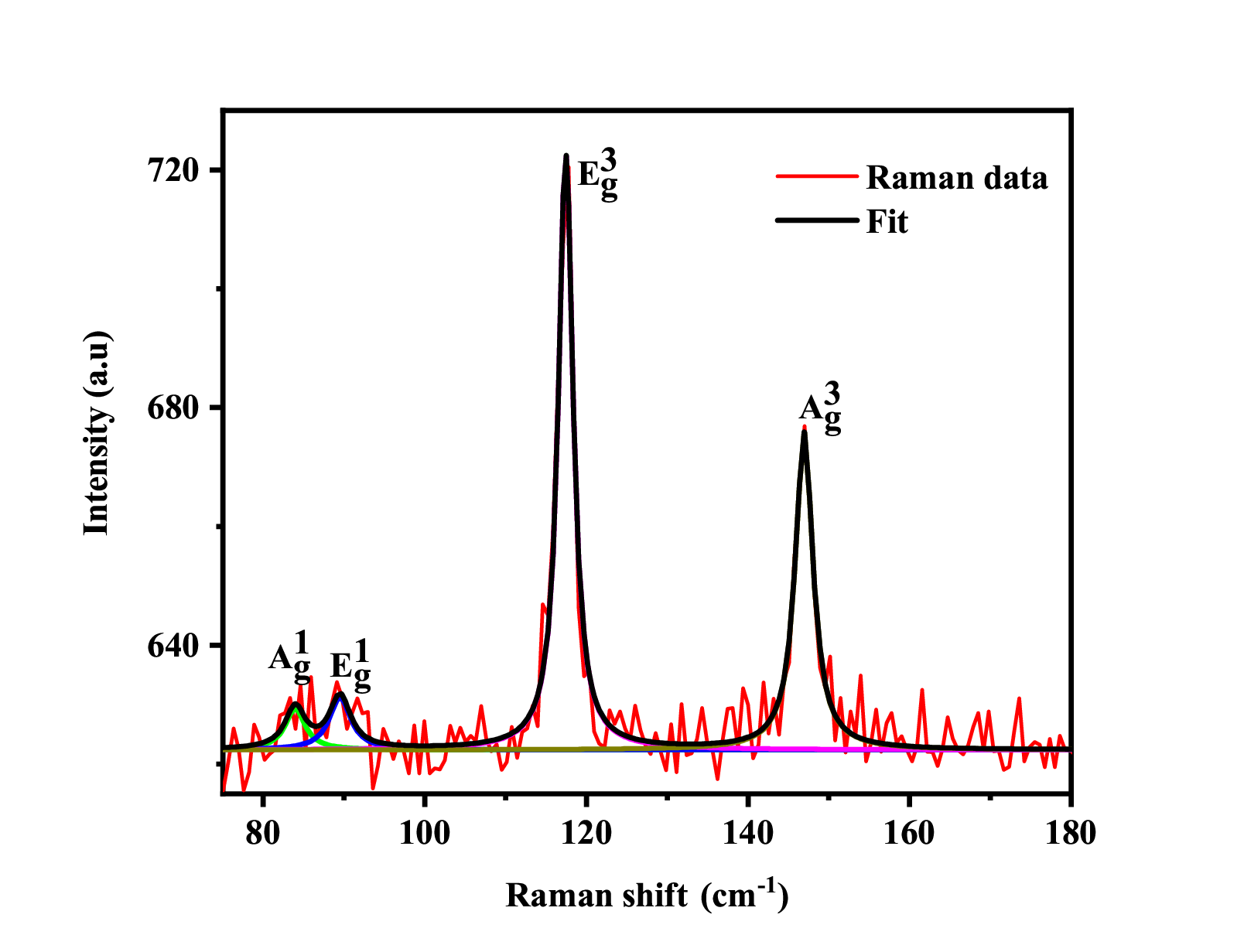}
\caption{Room temperature Raman Spectroscopy: Raman data (red lines) of bulk CrSiTe$_{3}$ along with its fit. The corresponding multi-Lorentzian fit shown in the black line. Five Lorentzian functions, represented individually by the green, blue,  magenta and dark yellow solid lines. }\label{Raman}
\end{figure}

Raman spectroscopy (Horiba Jobin Yvon - HR800, 488 nm excitation) was used to record the Raman spectrum from the sample. Figure \ref{Raman}  shows the Raman spectrum of bulk CrSiTe$_{3}$ single crystals. 
Multi-Lorentzian fitting of the Raman spectrum revealed four distinct peaks at 84 cm$^{-1}$, 89.5 cm$^{-1}$, 117.45 cm$^{-1}$ and 146.9 cm$^{-1}$, which are assigned to the A$_{g}^{1}$, E$_{g}^{1}$, E$_{g}^{3}$ and A$_{g}^{3}$ modes, respectively. Observed peak positions and full width half maxima (see Tab. \ref{raman_fit}) values match those reported in prior work \cite{milosavljevic2018evidence, zhang2022hard}.

\begin{table}
\caption{Raman spectroscopy  of CrSiTe$_{3}$:The extracted parameters from the multi-Lorentzian fit to the  Raman spectrum  are tabulated here}\label{raman_fit}%
\begin{tabular}{@{}lll@{}}
\hline\\
Raman mode &Wave number (cm$^{-1}$) &		FWHM (cm$^{-1}$) \\
\hline\\
A$_{g}^{1}$ &84.0 $\pm$ 0.6 & 	3.0 $\pm$  0.9 \\

E$_{g}^{1}$ &89.5 $\pm$  0.5 & 	3.2 $\pm$  1.0 \\

E$_{g}^{3}$&117.4$\pm$  0.03 & 	2.2 $\pm$  0.1 \\

A$_{g}^{3}$  &146.9 $\pm$  0.07 & 	2.6 $\pm$  0.2 \\

\hline\\
\end{tabular}
\end{table}

\subsection{Magnetization of CrSiTe$_{3}$.}

\begin{figure}%
\includegraphics[width= 1\textwidth]{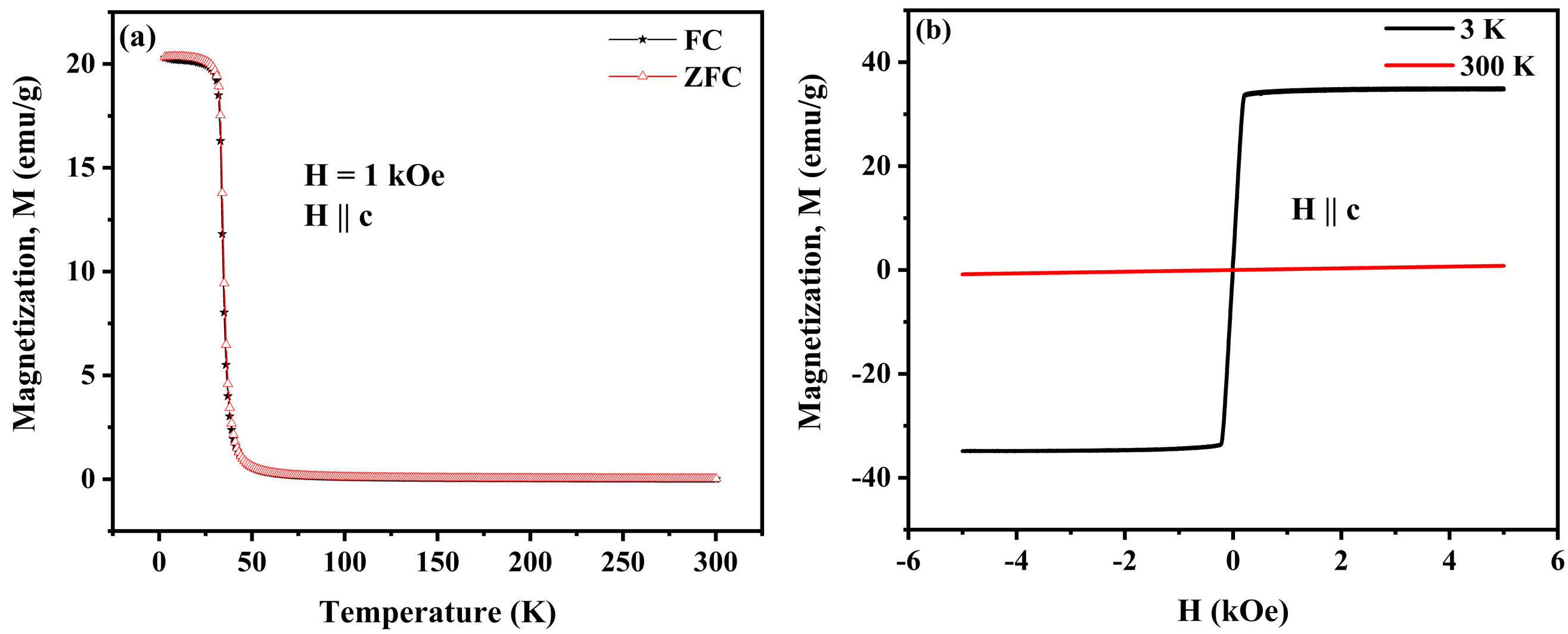}
\caption{Magnetization of CrSiTe$_{3}$ as a function of temperature and magnetic field: a) Magnetization (M) as a function of temperature measured under a magnetic field of 1 kOe applied along the c-axis, in field-cooled (black  line + closed star) and zero-field-cooled (red line + open triangle ) modes, respectively. (b) Magnetization as a function of magnetic field (H) along the c-axis, measured at 3 K (black solid line) and 300 K (red solid line), respectively. }\label{M_versus_H}
\end{figure}

Figure \ref{M_versus_H} (a) shows the temperature dependence of the magnetization (M) in field-cooled (red line) and zero-field-cooled (black line) modes under a magnetic field of 1 kOe applied along the c-axis. The temperature dependence of M reveals a ferromagnetic transition at a Curie temperature (T$_{c}$) of 33 K (refer Fig 2(b) of the main manuscript). Figure \ref{M_versus_H}(b) shows the magnetization as a function of magnetic field (H) along the c-axis at 3 K (black) and 300 K (red), respectively \cite{casto2015strong}.

\newpage

\section{Pump probe dynamics and data analysis.}
\begin{figure}%
\includegraphics[width= 1\textwidth]{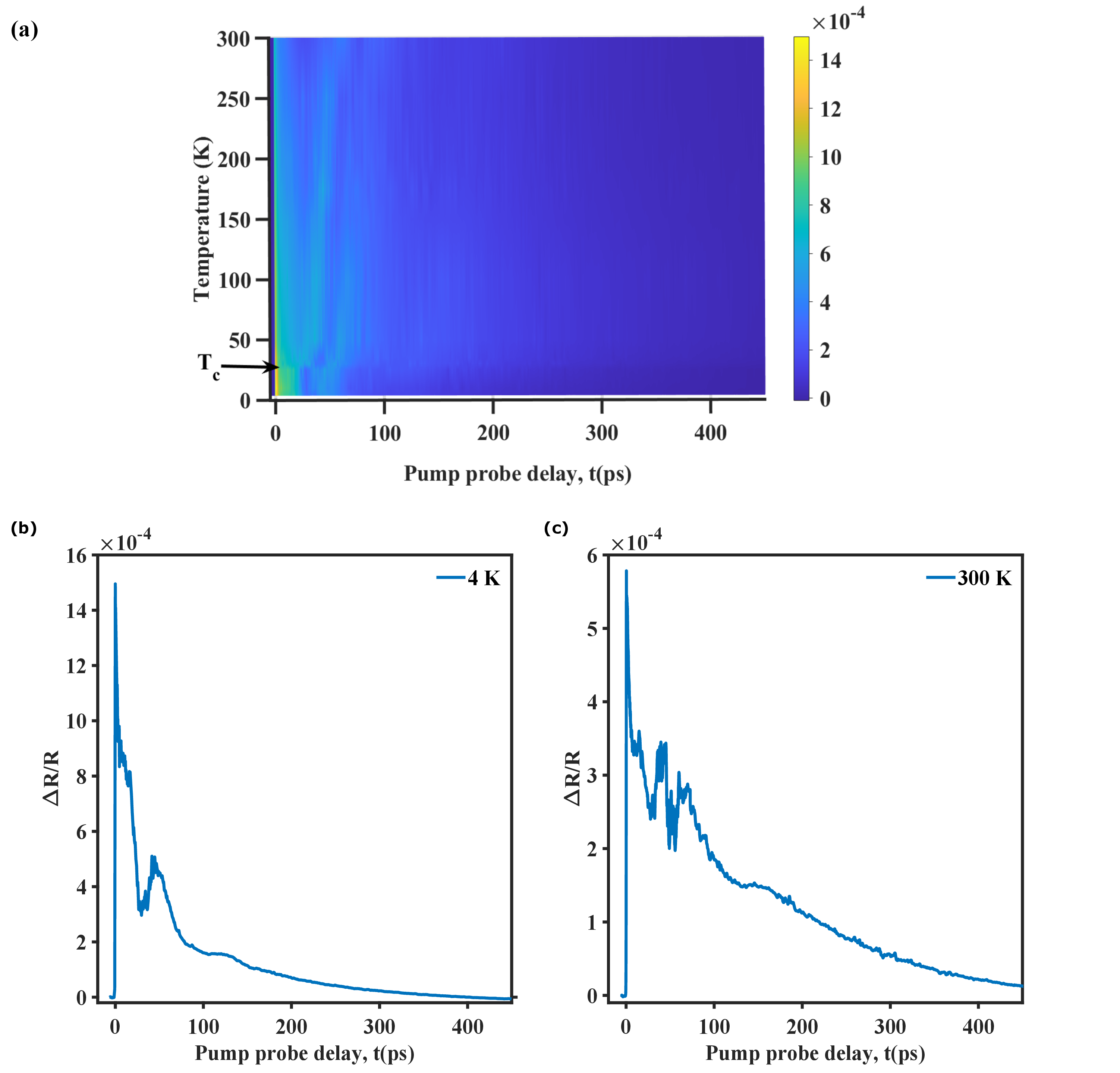}
\caption{Temperature evolution of pump-probe dynamics: (a) Two-dimensional color map of the temperature evolution of the transient differential reflectivity (DR) data of CrSiTe$_{3}$, shown up to 450 ps of the pump-probe delay at all measured temperatures from 4 K to 300 K. The black solid arrow indicates the Curie temperature (T$_{c}$). (b) and (c) shows the raw DR data at 4 K and 300 K, respectively. }\label{DR_dynamics}
\end{figure}

Figure \ref{DR_dynamics} (a) shows a 2D false-color map of the temperature evolution of the photoinduced transient differential reflectivity (DR) data of a single crystal of CrSiTe$_{3}$ pumped at 650 nm with a fluence of 1.77 mJ/cm$^{2}$. As indicated by the black arrow in Fig. \ref{DR_dynamics}(a), a noticeable change in the raw DR/R data can be seen across the Curie temperature (T$_{c}$). This change indicates that the transient reflectivity data is indeed affected by the magnetic phase transition, which is the first signature of magneto-elastic coupling in CrSiTe$_{3}$. For clarity, the data at 4 K and 300 K are shown in Fig. \ref{DR_dynamics}(b,c). The photoinduced transient reflectivity data has three main features at all temperatures: a positive peak at zero pump-probe delay, followed by strain-induced oscillation on top of the exponentially decaying electronic part.

\begin{figure}%
\includegraphics[width= 0.8\textwidth]{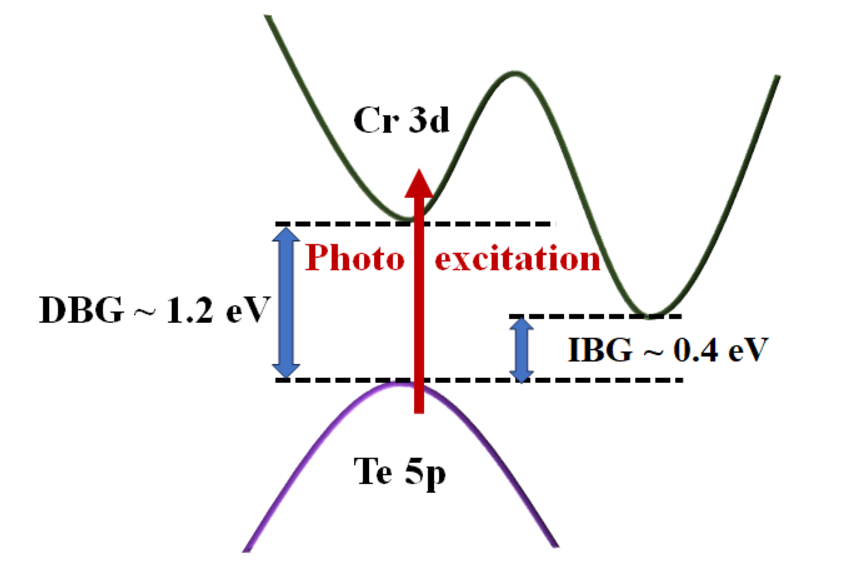}
\caption{Schematic band diagram of CrSiTe$_{3}$: 
 The valance band is mainly composed of Te 5p levels and conduction band is mainly composed of Cr 3d levels. The direct (DBG) and in-direct band gap (IBG) is $\sim$ 1.2 eV and $\sim$ 0.4 eV, respectively. Photo-excitation using 1.9 eV is indicated in solid red arrow.}\label{Band_schematic}
\end{figure}

The DR response upon pump excitation in semiconductors is caused by the change of the complex refractive index of the sample associated with optical transitions\cite{weber_APL2015,jnawali2021band} and results in the positive sign. CrSiTe$_{3}$ has an indirect band gap of 0.4 eV and a direct band gap of 1.2 eV. The valence band of CrSiTe$_{3}$ is primarily composed of Te 5p levels and the conduction band is mainly composed of Cr 3d levels (refer to Fig. \ref{Band_schematic}) \cite{kang2019effect, kang2023field}. The central photon energy of the pump ($\sim$ 1.9 eV) and probe ($\sim$ 1.55 eV) pulses in our experiment is within  the direct band gap of CrSiTe$_{3}$. As a result, both the pump and the probe pulses can excite electrons from the Te p levels to the Cr d levels \cite{ron2020ultrafast}. This leads to a reduction in the absorption of the probe pulse in the presence of a strong pump pulse, which is known as the state filling effect.

\begin{figure}%
\includegraphics[width= 1\textwidth]{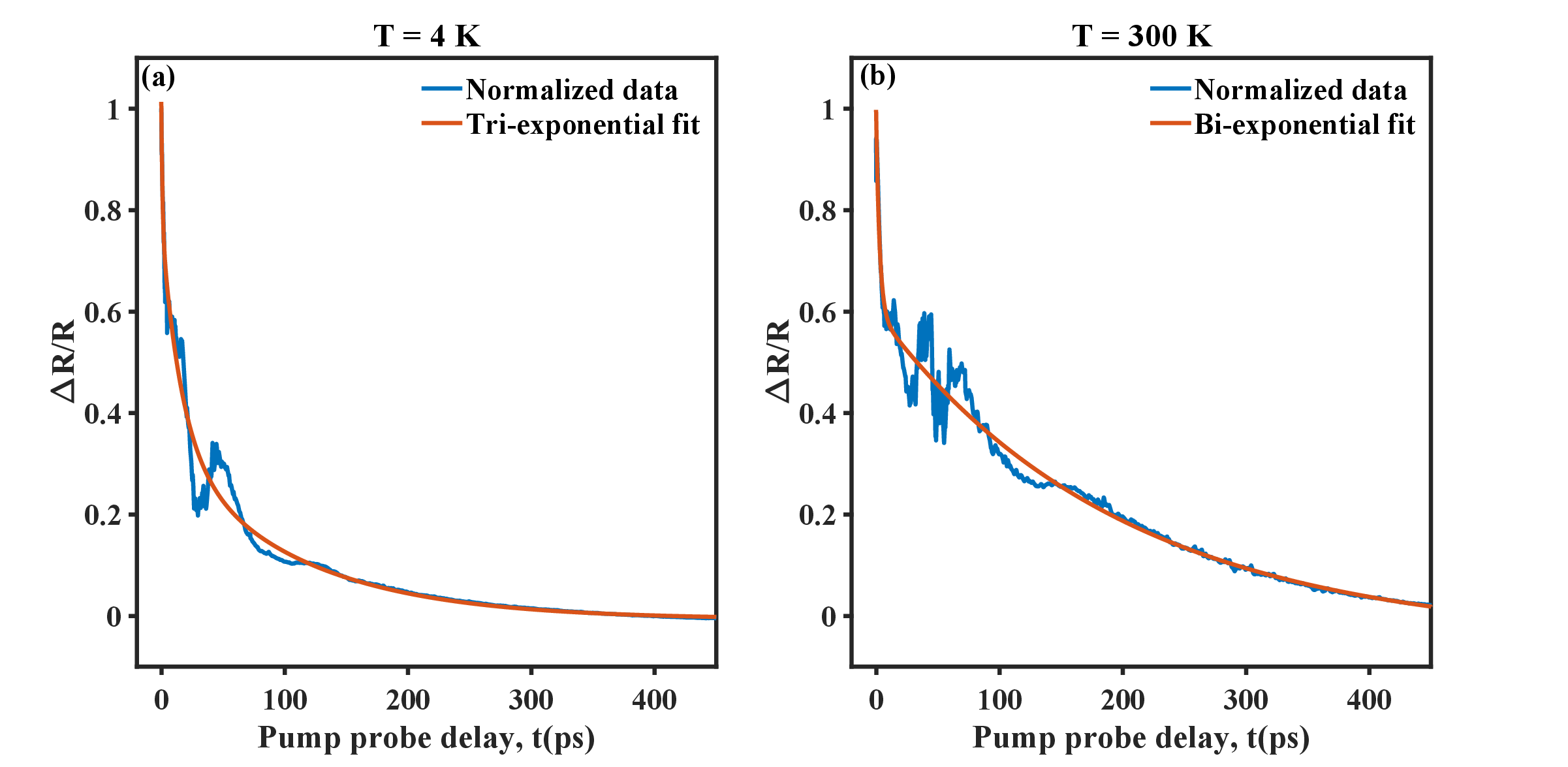}
\caption{Raw transient differential reflectivity data and its multi-exponential fit: Normalized transient differential reflectivity data (blue solid line) of CrSiTe$_{3}$ single crystal and their fit to tri exponential (orange solid line) and Bi-exponential functions (orange solid line) at (a) 4 K and (b) 300K, respectively presented until the pump probe delay time (t) of 450 ps. The data is normalized with respect to its peak value. }\label{Multi_exp_fit}
\end{figure}

To understand the carrier dynamics in CrSiTe$_{3}$, the exponentially decaying electronic background of the DR data was fitted with a multi-exponential decay function.  At all the temperatures, the DR data was normalized to its peak value and cut at 0 ps and fitted with a multi-exponential function up to the pump probe delay of 450 ps to extract the relaxation timescales and the decay coefficients.  Interestingly, below 35 K (including 35 K), the data could be successfully fitted to a minimum sum of three exponential functions (shown in Fig. \ref{Multi_exp_fit} (a)) with a constant background A$_{0}$, given by 

\begin{equation}
\Delta R/R = \sum_{i=1}^{3} A_{i}exp(-t/\tau_{i}) + A_{0}\label{tri_exp} 
\end{equation}

Whereas, above 35 K (from 50K data) the data could be successfully fitted to minimum sum of bi-exponential functions (shown in Fig. \ref{Multi_exp_fit} (b)) with a constant background, 

\begin{equation}
\Delta R/R = \sum_{i=1}^{2} A_{i}exp(-t/\tau_{i}) + A_{0}\label{Bi_exp} 
\end{equation}

where A's are the decay coefficients and $\tau$'s are the relaxation timescales. 

\begin{figure}%
\includegraphics[width= 1\textwidth]{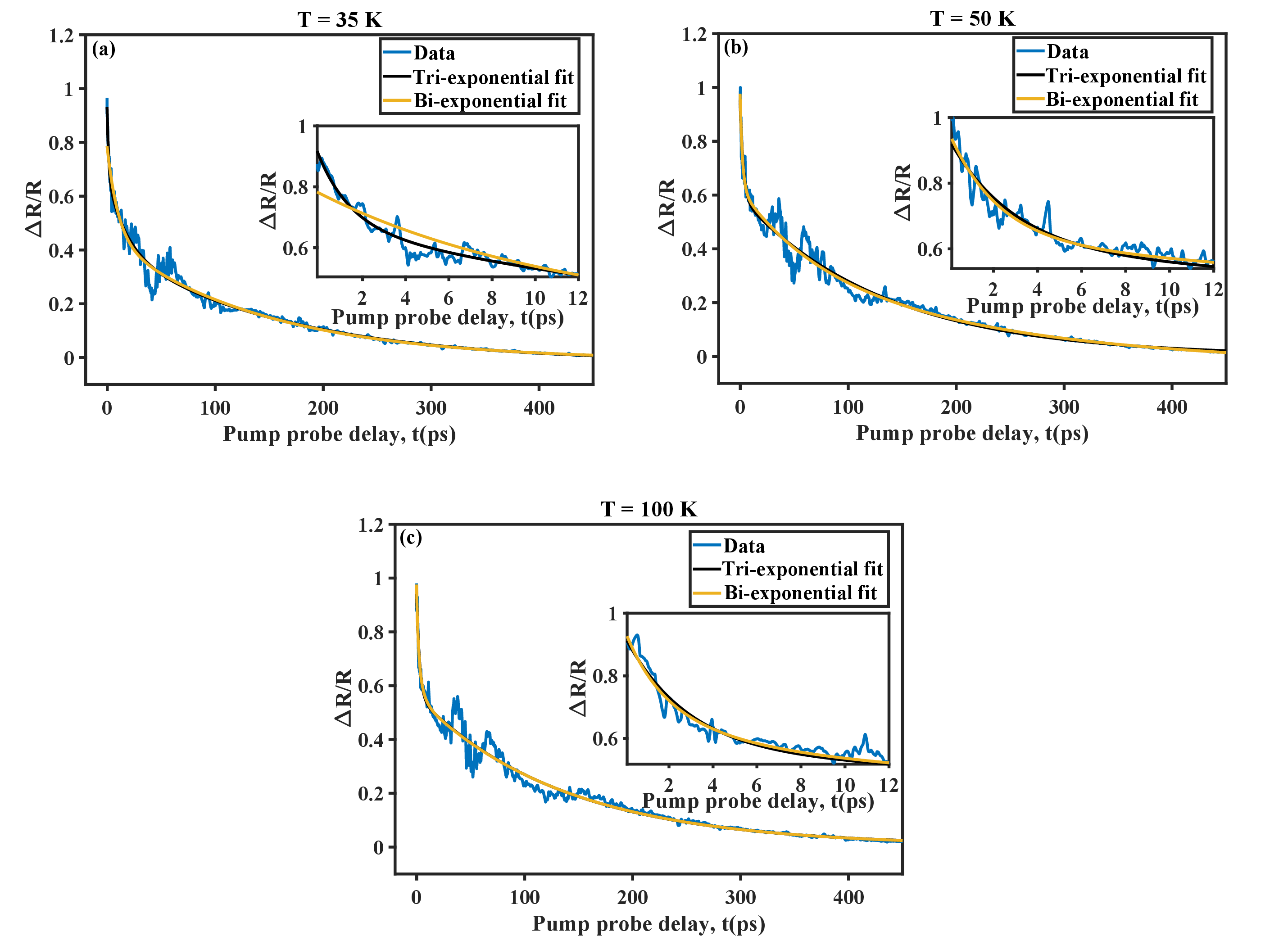}
\caption{Comparison of Bi and Tri-exponential fits to the DR data:  The transient differential reflectivity data (blue solid line) normalized with respect to its peak value and cut at 0 ps of the pump probe delay is fitted with both minimum sum of two exponentials (yellow solid line) and minimum sum of three exponential functions (black solid line) for comparison at (a) 35 K, (b) 50 K and (c) 100 K respectively. For clarity, the data and the respective fits are shown up to 12 ps of the pump probe delay time in the insets of each figures. }\label{fitting_comparison}
\end{figure}

To justify why at least three exponentials are needed to fit the DR data below 35 K, the data at 35 K was fit with both two- and three-exponential functions (Fig. \ref{fitting_comparison}).  From the figure \ref{fitting_comparison} (a) and its inset, it is clear graphically that the tri-exponential function (black solid line) fits the data well, while the bi-exponential function (yellow solid line) deviates from the experimental DR data (blue solid line). Therefore, at least three exponentials are needed to fit the DR data below 35 K.
At 50 K and higher temperatures, the data could be fitted with both two- and three-exponential functions ( refer to Fig. \ref{fitting_comparison} (b) and (c), where the data and the fit at 50 K and 100 k are shown for clarity). The fits are similar in both cases (refer to inset of Fig. \ref{fitting_comparison} b and c), and the goodness of fit (R$^{2}$ value) is nearly identical: 0.977 and 0.978 for bi- and tri-exponential fits at 50 K, and 0.974  for both  bi- and tri-exponential fits at 100 K. Therefore, we prefer to fit the data at and above 50 K with two exponentials, since it has minimum number of exponentials.

\newpage
\section{Decay coefficients and their temperature dependence.}

\begin{figure}%
\includegraphics[width= 1\textwidth]{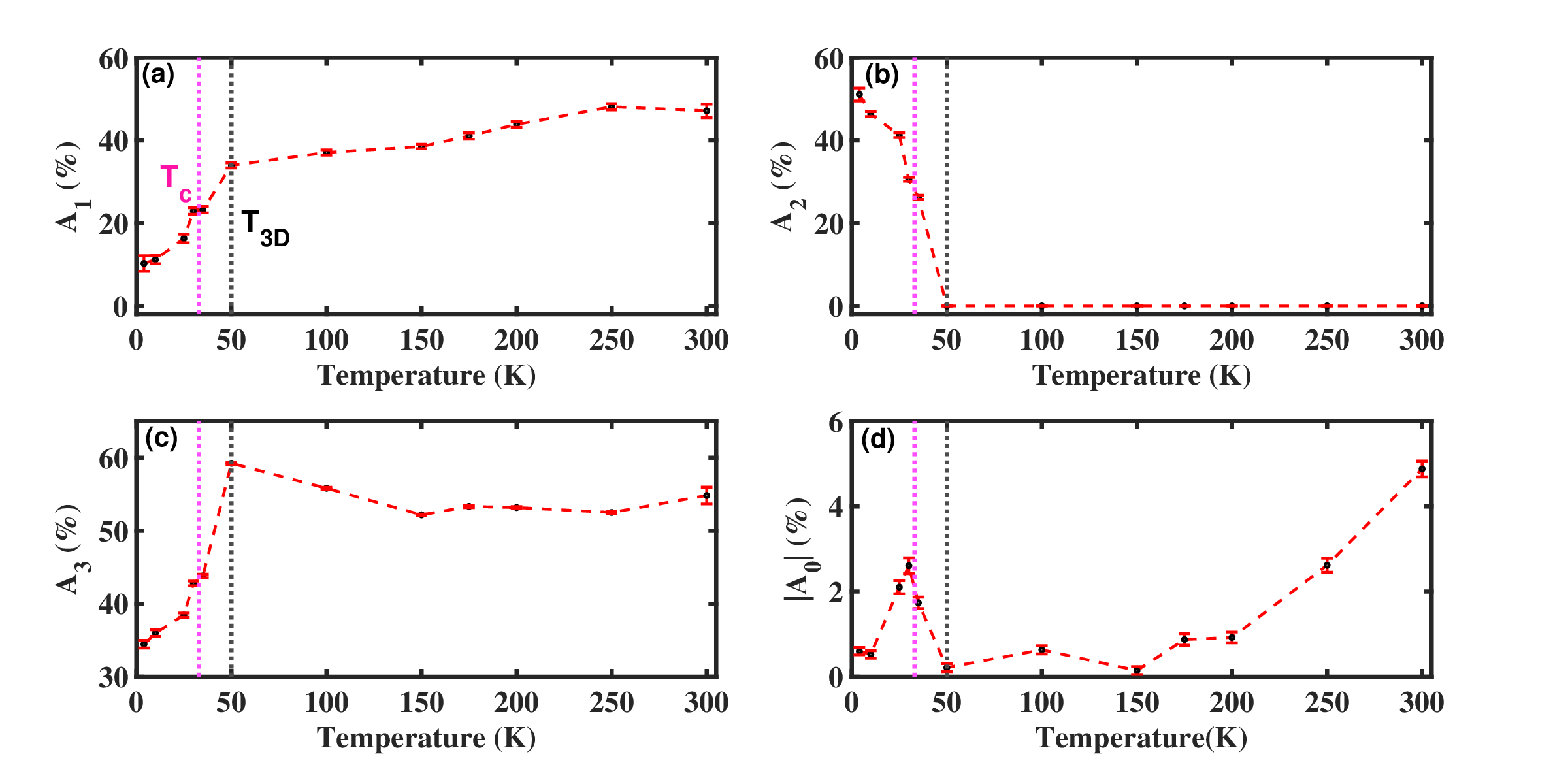}
\caption{Temperature dependent decay coefficients:  The variation of the decay coefficients (a) A$_{1}$, (b) A$_{2}$,  (c) A$_{3}$  and (d) absolute value of A$_{0}$ as a function of the temperature. The error bars are determined using orthogonal distance regression (ODR) approach.  The red dashed lines are guide to eye. The magenta and black vertical dotted line indicates the Curie temperature (T$_{c}$) and the temperature (T$_{3D}$) at which the onset of LRMO fluctuations begin. }\label{Decay_coefficients}
\end{figure}
The relaxation timescales and decay coefficients of the transient differential reflectivity (DR) data are extracted from a multi-exponential fit. The relaxation timescales and their temperature dependence are discussed in the main manuscript (refer to Fig. 2 (b) of the manuscript), while the decay coefficients and their temperature dependence are shown in Fig. \ref{Decay_coefficients}.  A$_{1}$ varies from $\sim$ 15 \% to $\sim$ 40 \% in the temperature range  4 K to 50 K.  Through-out this temperature range A$_{1}$ increases with the increasing temperature. Also, the relaxation timescale, $\tau_{1}$ (1-3 ps) associated with this coefficient suggest this process is associated with the electron-lattice - spin thermalization (see main manuscript) \cite{jnawali2021band}. A$_{2}$ is only present below 50 K and decreases from $\sim$ 50 \% to $\sim$ 20 \% (from 4 K to 35 K) as the temperature increases. Therefore this processes is attributed to the interaction of   lattice with interlayer spin-spin fluctuations.  A$_{3}$ increases from $\sim$ 20 \% to $\sim$ 60 \% as the temperature increases from 4 K to 50 K, and then remains constant with the temperature above 50 K. This process is associated with the lattice interaction with the intralayer spin-spin fluctuations. A$_{0}$ is the coefficient associated with the long decay process and it varies from $\sim$ 2 to $\sim$ 4 \%. The similar relaxation timescales and decay coefficients have been seen in the nano-sheet of CrSiTe$_{3}$ \cite{jnawali2021band}. These results suggest that all the decay coefficients undergo a drastic change at the T$_{3D}$ temperature, which is the onset of the long-range magnetic order (LRMO) fluctuations. 

\newpage

\section{Picosecond strain pulse: Origin of strain pulse and the effect of magnetic ordering on the strain pulse.}

\begin{figure}%
\includegraphics[width= 1\textwidth]{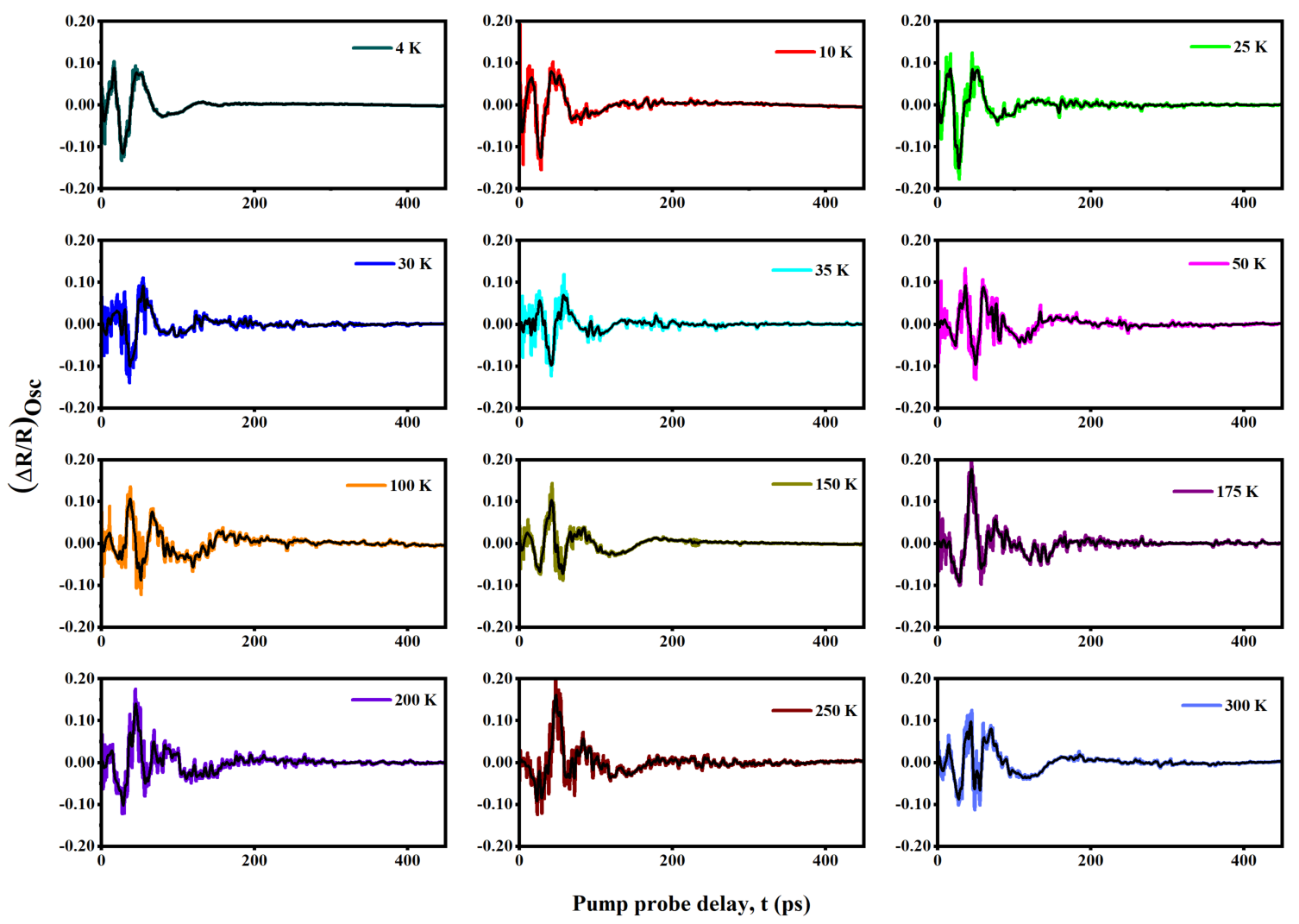}
\caption{Strain pulses at various temperatures: This figure shows strain pulses extracted from transient differential reflectivity data for CrSiTe$_{3}$ at various temperatures, represented by different colored solid lines. The smoothened data at each temperature is shown in the black solid lines}\label{oscillatuions_all_T}
\end{figure}

\begin{figure}%
\includegraphics[width= 0.9\textwidth]{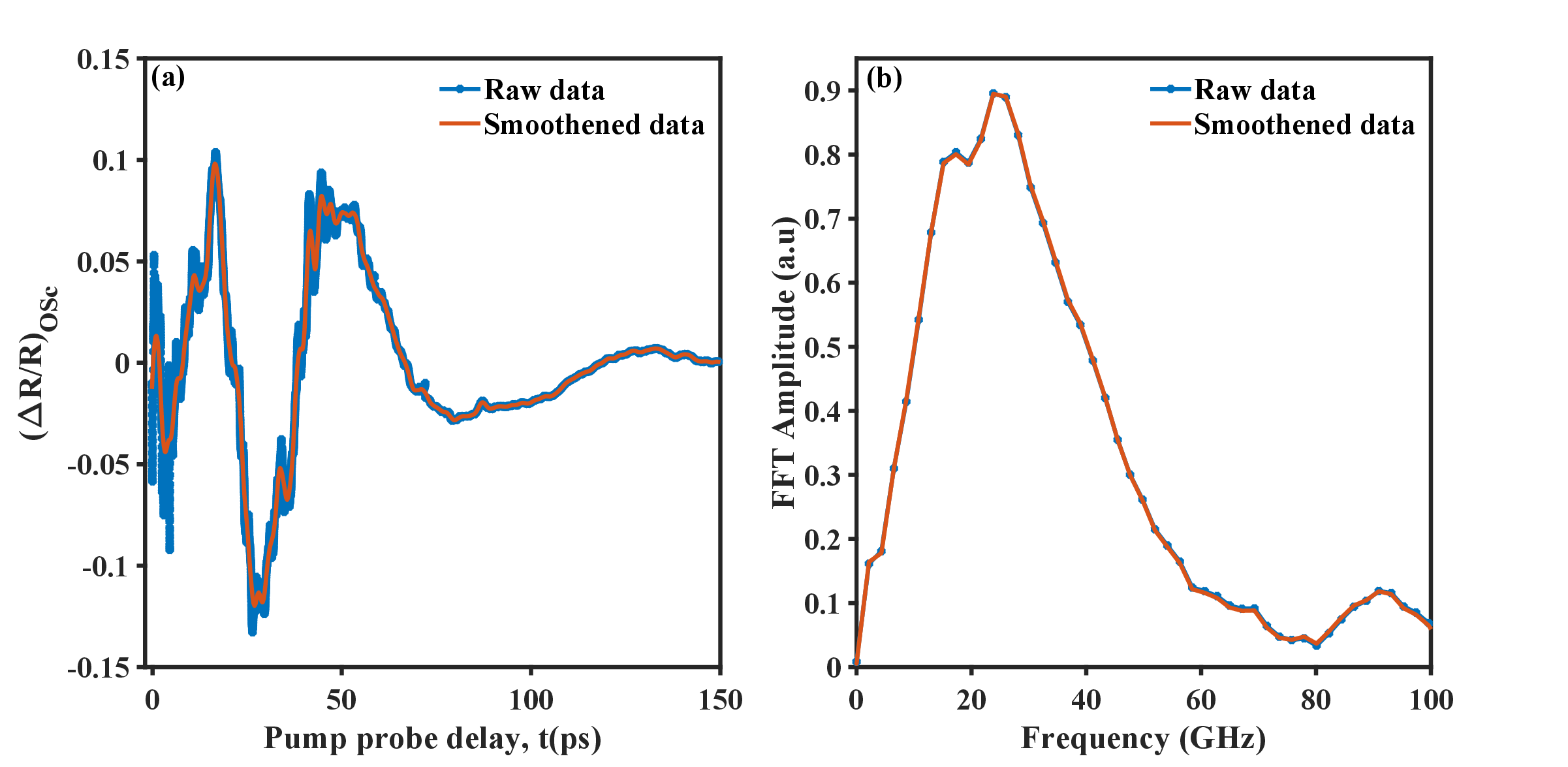}
\caption{Data processing: (a) Time-domain strain pulse of CrSiTe$_{3}$ at 4K. The raw data is shown in blue squares with line, and the smoothened data is shown in orange solid line. The corresponding FFT spectra are shown in panel (b)}\label{Smoothening_oscillations}
\end{figure}

The interesting feature of the DR data is the strain pulse.  We extracted the oscillatory component by subtracting the multi-exponential fits at each temperature. Figure \ref{oscillatuions_all_T} shows the temperature evolution of DR oscillations for all temperatures and pump-probe delay times up to $\sim$ 450 ps. The variation of the strain pulse as a function of temperature is discussed in the main manuscript, therefore this section focuses on the origin of the strain pulse. For clarity, smoothened data of strain pulse at various temperatures is shown the Fig 2 (d) of the main manuscript. The smoothening is done by using the smoothening function ‘Loess’ with the x-axis increment 0.01 using MATLAB software. Figure \ref{Smoothening_oscillations} (a) here compares the raw strain pulse data (various colored lines) and its smoothened data (black line) in the time domain. The corresponding FFT spectra of raw and smoothened data are shown in Fig \ref{Smoothening_oscillations} (b). It is important to note that the smoothening  does not alter the data, which is evident from the FFT spectrum of both raw and smoothened data shown in Fig. \ref{Smoothening_oscillations} (b). 

The  strain pulses in Fig. \ref{Smoothening_oscillations} (a) resemble a single-cycle pulse rather than sinusoidal oscillations. These  are caused by lattice strain which in turn is associated with lattice stress. Several mechanisms can contribute to lattice stress in materials, including thermal expansion (TE) stress \cite{thomsen1984coherent}, deformation potential (DP) interaction\cite{wu2007femtosecond}, inverse piezoelectric interaction\cite{babilotte2010picosecond}, and magnetic stresses\cite{schmising2008ultrafast, von2020unconventional, mattern2023towards}. TE stress arises from the transfer of excess energy from photoexcited carriers to the lattice via electron-phonon interaction as they relax to the band edge, causing a rapid increase in lattice temperature. DP interaction occurs when photo-excitation of electrons from the valence band to the conduction band disrupts lattice equilibrium, leading to lattice deformation. This deformation alters the semiconductor band structure and induces electronic stress associated with  free carriers in conduction band. Inverse piezoelectric interaction arises when photoexcited charge carriers screen the intrinsic built-in electric field in non-centro-symmetric materials, modifying the lattice equilibrium state via the inverse piezoelectric effect and generating inverse piezoelectric stress (IP). Since CrSiTe$_{3}$ is a centro-symmetric crystal, the piezoelectric effect is negligible. The  presence of electron-phonon thermalization (decay coefficient A$_{1}$ $\sim$ 40\%) in the electronic background of DR and at typical photoexcited carrier density N $\sim$ 4.2$\times$ 10$^{20}$ cm$^{-3}$ (estimated using N = $\frac{F\alpha R}{E}$, where F is the fluence in mJ/cm$^{2}$, $\alpha$ is the absorption coefficient in cm$^{-1}$, R is the reflectivity, and E is the pump photon energy), both TE and DP processes dominantly contribute to the strain pulse in this material in the paramagnetic phase. In the ferromagnetic compound CrSiTe$_{3}$, the magnetic order arises from super-exchange interactions between chromium (Cr) metal atoms mediated by tellurium (Te) ligands. A recent study\cite{ron2020ultrafast} suggests that photo-excitation-induced charge transfer from ligands to metal ions can significantly increase the super-exchange interaction energy. This strengthened interaction, in turn, will introduce additional magnetic stress through the magnetostrictive mechanism in the ferromagnetic state. Therefore in the ferromagnetic phase, the strain arise from all three processes, which is given by\cite{mattern2023towards}

\begin{equation}
\eta_{Total}(x, t) = \eta_{TE}(x, t) + \eta_{DP}(x, t)+ \eta_{Magnetic} (x, t)  \label{Total_strain}  
\end{equation} 
where $\eta_{TE}(x, t)$ , $\eta_{DP}(x, t)$, $\eta_{Magnetic}(x, t)$ are the strain due to TE, DP and Magnetic origins. 

Magnetic materials often thus exhibit a competition between contrasting forces: expansive stresses arising from elastic processes (TE and DP) and a contractive stress induced by magnetic order. This interplay leads to the emergence of intricate spatio-temporal strain pulses, a phenomenon observed in materials like dysprosium \cite{von2020unconventional, mattern2023towards}, FePt thin films\cite{reid2018beyond, von2020spin}, and SrRuO$_{3}$ nano-layers in a SrRuO$_{3}$/SrTiO$_{3}$ super-lattices \cite{schmising2008ultrafast}. Alexander von Reppert et al.\cite{von2020unconventional} have  explained this complex interaction by modeling the spin system as an energy reservoir with a saturation limit. This reservoir generates substantial contractive stress on ultrafast timescales, explaining the observed intricate strain response. Additionally, their model enabled the estimation of the time- and space-dependent magnetic stress.

\begin{figure}%
\includegraphics[width= 0.9\textwidth]{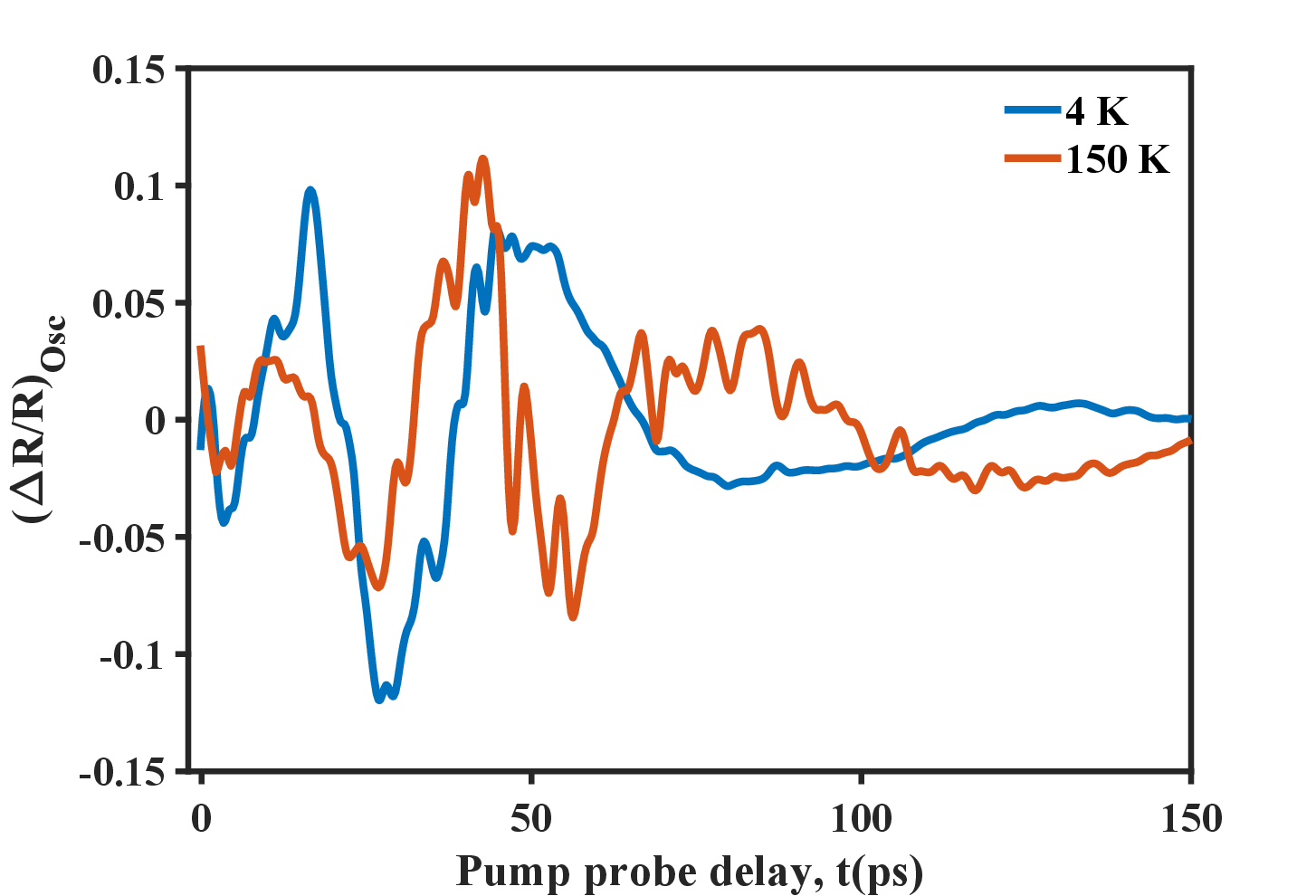}
\caption{Comparison of strain pulse in ferromagnetic and paramagnetic phases: The figure shows the smoothened time-domain strain pulse for CrSiTe$_{3}$ at 4 K (blue solid line) and 150 K (orange solid line) for the first 150 ps after pump excitation}\label{FM_PM_comparison}
\end{figure}

The interplay between various strain generation processes enable  us to capture the each distinct step of  the magnetic dimensional crossover using picosecond strain pulses. The main manuscript discusses the effect of magnetic dimensional crossover on the strain pulses. For example,  Fig. \ref{FM_PM_comparison} shows the strain pulse in the ferromagnetic phase (4 K) and in the paramagnetic phase (150 K). For clarity, only data at two temperatures are shown here. As evident from Fig. \ref{FM_PM_comparison}, the strain pulse at 4 K is inverted compared to the strain pulse at 150 K. This indicates that the contractive magnetic stress in the FM phase dominates over the expansive elastic stress.

\begin{figure}%
\includegraphics[width= 1\textwidth]{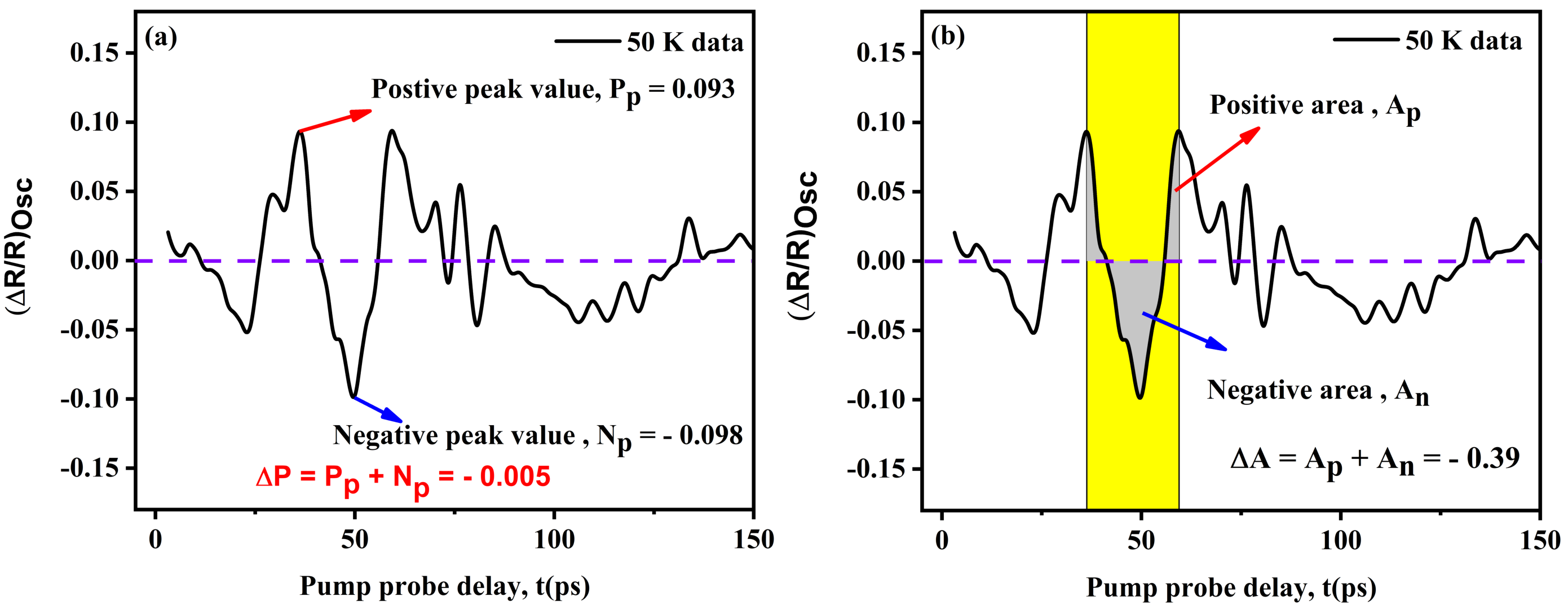}
\caption{Extraction of ($\Delta P$) and ($\Delta A$): (a) The peak-to-peak difference ($\Delta P$) of the central peak of strain pulse is shown here. $P_{p}$ represents the positive peak value (indicated in the solid red arrow) and  $N_{p}$ represents the negative peak value (indicated in the solid blue arrow) of the central peak, respectively. (b) The area difference ($\Delta A$) between  the positive area and negative area of the central part of the  strain pulse is shown here. $A_{p}$ represents the positive area (indicated in the solid red arrow) of the main central part of the strain pulse  and  $A_{n}$ represents the negative area of the central part of the strain pulse (indicated in the solid blue arrow). Both the areas are  shown in shaded grey colour. The purple dashed line marks y = 0. }\label{peak_area_procedure}
\end{figure}

To quantify the interaction between the processes, we calculated the peak-to-peak difference ($\Delta P$) and the area difference ($\Delta A$) of the positive and negative portions of the central part of the strain pulses. $\Delta P$ was obtained by summing the positive peak value (P$_{p}$) and the negative peak value (P$_{n}$) of the central peak. To account for potential fluctuations in the peak values, we also calculated $\Delta A$ by adding the positive area (A$_{p}$) and negative area (A$_{n}$) of the central part of the strain pulse. The extraction procedure for both $\Delta P$ and $\Delta A$ is illustrated in Fig. \ref{peak_area_procedure} for data at 50K. The same procedure was repeated for all temperatures for consistency.

\section{Picosecond strain pulse: frequency domain spectrum.}

\subsection{Fast Fourier Transform (FFT)}

\begin{figure}%
\includegraphics[width= 1\textwidth]{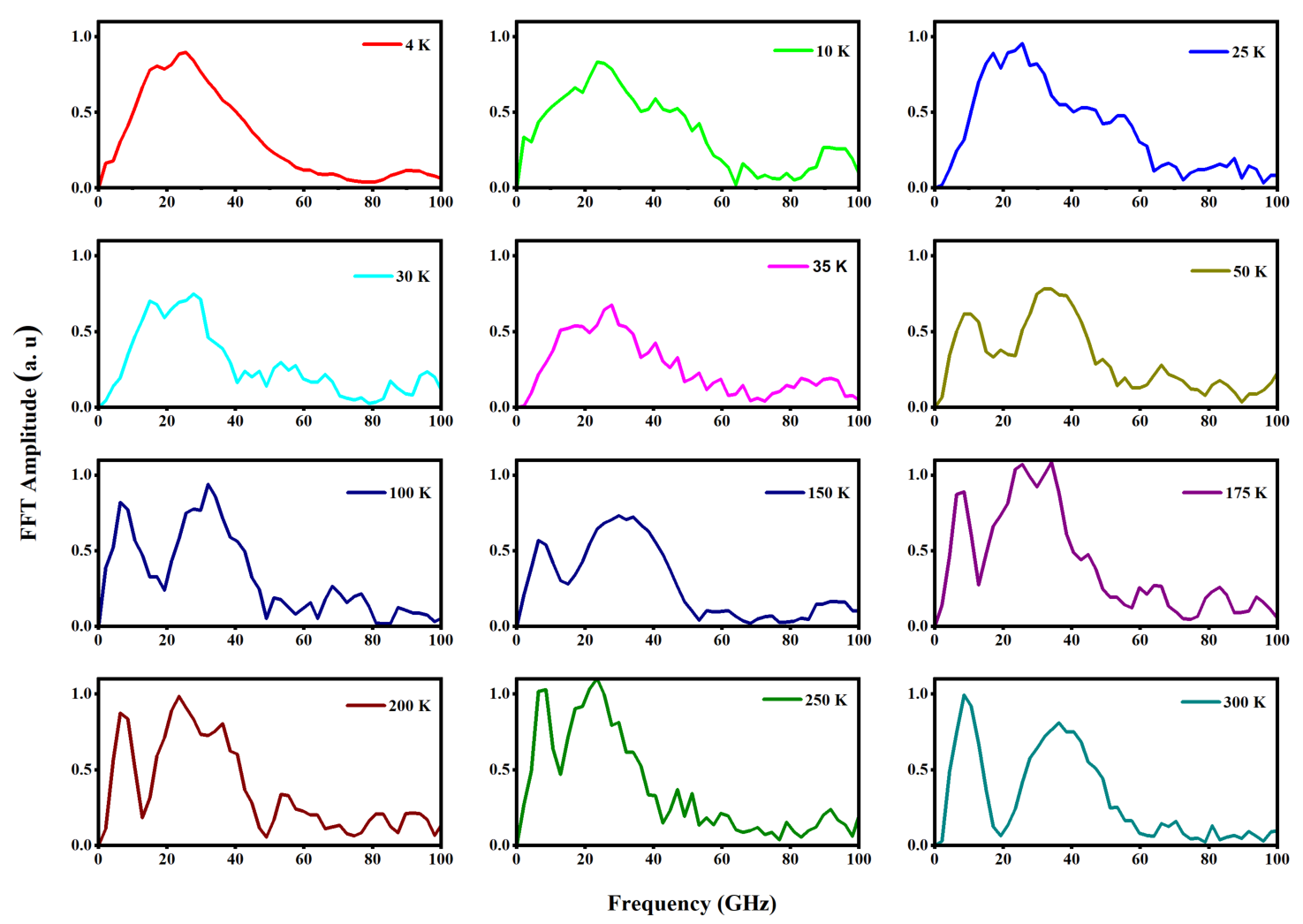}
\caption{The frequency domain spectrum at various temperatures: This figure shows the FFT amplitude as a function of frequency  obtained by fast Fourier transform of time domain strain pulse data for CrSiTe$_{3}$ at various temperatures, represented by different colored solid lines.}\label{T_dependent_FFT}
\end{figure}

The frequency domain spectrum of the strain pulse is obtained  using the fast Fourier transform (FFT). The FFT of the time domain strain pulses at different temperatures are shown in Fig. \ref{T_dependent_FFT}. The main manuscript discusses the variation of the frequency components of the strain pulse in detail.

\newpage

\subsection{Continuous Wavelet Transform (CWT) }
To track the time-dependent behavior of strain pulses, we employed a continuous-wavelet transform (CWT) approach \cite{torrence1998practical, hase2003birth, kamaraju2010temperature}. Leveraging the Gabor mother wavelet, a modified MATLAB code based on a Gaussian function was utilized to achieve this

\begin{equation}
\psi(t/s) = \pi^{-1/4} \left(\frac{1}{ps}\right)^{1/2}  exp\left(-\frac{t^{2}}{2s^{2}p^{2}} + \sqrt{-1} \frac{t}{s}\right), \label{psi_CWT}  
\end{equation} 

the scaling factor $\rq s \rq$ corresponds to the inverse of the frequency, while $\rq p \rq $ is a constant defined as $\pi (2/ln(2))^{1/2}$.  This section outlines the process of calculating the CWT. The wavelet transform of a given time signal, x(t), is represented by

\begin{equation}
CWT (\tau, s) = \frac{1}{\sqrt{|s|}} \int x(t) \psi^{*}((t-\tau)/s)  dt\label{CWT_amplitude}  
\end{equation} 

To initiate the wavelet transform, a starting scale ($s_{start}$) is selected, corresponding to a frequency higher than the signal's highest frequency (determined by FFT). This initial wavelet, denoted as $\psi_{start}(t/s_{start})$, is a compressed version of the signal in the time domain. The cross-correlation of $\psi_{start}(t)$ with x(t) is calculated using equation \ref{CWT_amplitude} with $\tau$ = 0 ($\tau$ is the transnational time). The magnitude of this cross-correlation reflects the similarity between the frequency components of x(t) and $\psi_{start}(t)$. This process is repeated by translating $\psi_{start}(t)$ in the time domain ($\tau$), yielding wavelet coefficients for a specific scale (s) and a range of $\tau$ values. The procedure is then iterated for progressively higher scales (s). This results in a collection of wavelet coefficients for x(t) across the time-scale plane. This plane is subsequently transformed into the time-frequency plane. The wavelet transform of a time-domain signal produces a three-dimensional (3D) plot of wavelet coefficients versus frequency and time.

\begin{figure}%
\includegraphics[width= 1\textwidth]{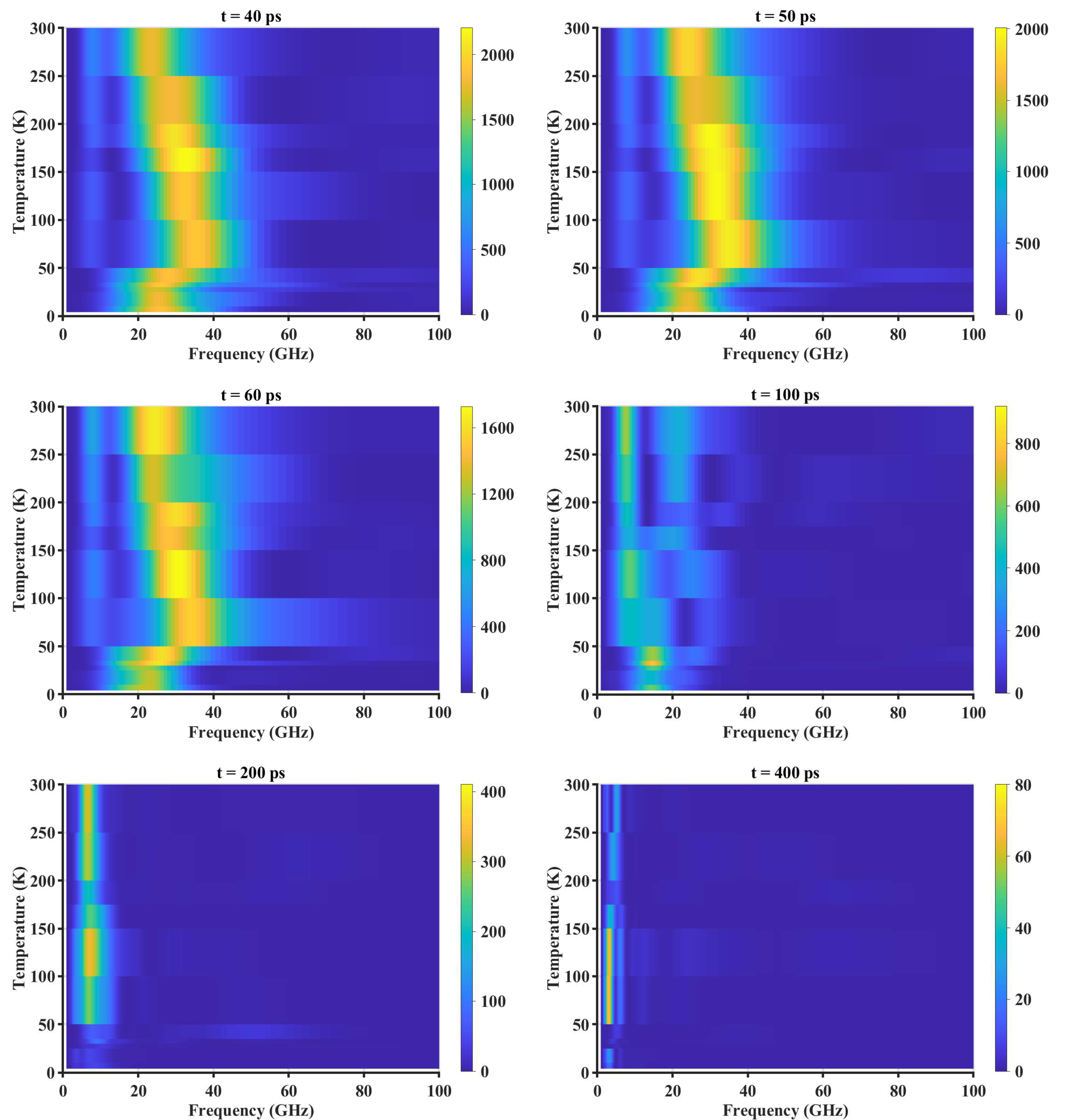}
\caption{CWT spectra at various pump probe delay times: The evolution of the CWT amplitude  with temperature at different pump-probe delay times. For the clarity, CWT spectra at the pump probe delay time 40 ps, 50 ps, 60 ps, 100 ps, 200 ps and 400 ps are shown here}\label{time_dependent_CWT}
\end{figure}

Figure \ref{time_dependent_CWT} illustrates the evolution of the CWT spectrum (frequency spectrum of the strain pulse) as a function of  temperature at various pump-probe delay times obtained from continuous wavelet transform. From the Fig. \ref{time_dependent_CWT} it is evident that there are two parts in the spectrum: high frequency acoustic part (HFAP) and low frequency acoustic part (LFAP).   Both the modes appear within a few picoseconds of pump excitation (data not shown here). The amplitude of HFAP peaks at $\sim$ 50 ps and decays within $\sim$ 150 ps, while the LFAP peaks at $\sim$ 80 ps  and then decays within $\sim$ 400 ps, indicates its long-lived nature. In the entire time scale, the nature of the frequency spectrum remains unchanged at t $>$ 35 ps. The variation of the acoustic mode frequency with the temperature (T $>$ T$_{3D}$) can be described using  the anharmonic decay model given by \cite{balkanski1983anharmonic, orbach1964attenuation, klemens1967decay, baumgartner1981spontaneous, tamura1985spontaneous}, 

\begin{equation}
\omega_{i}(T) =   \omega_{i}(0) + (\Delta \omega_{i})_{anh}(T) \label{p1} 
\end{equation}

where, $\omega_{i}(0)$ represents  the frequency of the $i^{th}$ phonon mode at T = 0 K  and $(\Delta \omega_{i})_{anh}(T)$ is the shift in the frequency of the $i^{th}$ phonon mode due to the anharmonic phonon-phonon coupling, which is given by 

\begin{equation}
(\Delta \omega_{i})_{anh}(T)  =  C \left[1 + \frac{2}{exp(\frac{\hbar \omega_{i}(0)}{2k_{B}T})-1} \right]  \label{p2} 
\end{equation}

where, C is the constant which signifies the strength of the decay process. $\hbar$ and $k_{B}$ are the reduced Planck's constant and Boltzmann constant, respectively. This model incorporates cubic anharmonicity, which describes how a  high-frequency   phonons can decay into two low frequency  phonon modes with equal frequencies \cite{orbach1964attenuation, klemens1967decay, baumgartner1981spontaneous, tamura1985spontaneous}. The observed frequency shifts of LFAP and HFAP are thus fitted with the anharmonic model and the fit is shown in Fig. 3(b) of the main manuscript. The anharmonic model successfully captures the frequency shifts observed in both the  LFAP and the HFAP. The fitting  reveals that the zero-frequency values of LFAP and HFAP are $\omega_{LFAM}(0)$ = 198 GHz and $\omega_{HFAM}(0)$ = 225 GHz, respectively. Notably, the coupling constants $C_{LFAP}$ and $C_{HFAP}$ are determined to be - 17 $\times 10^{-4}$ cm$^{-1}$ and - 8000 $\times 10^{-4}$ cm$^{-1}$, respectively. This substantial difference in the magnitudes of $C_{LFAM}$ and $C_{HFAM}$ indicates that the anharmonic coupling is significantly weaker for LFAP compared to HFAP. This is in line with the fact that the high frequency phonons decay to low frequency phonons and hence the decay rate is higher for HFAP; whereas for LFAP, the decay rate is slower as they are already in lower in energy/frequency.

\begin{figure}%
\includegraphics[width= 1\textwidth]{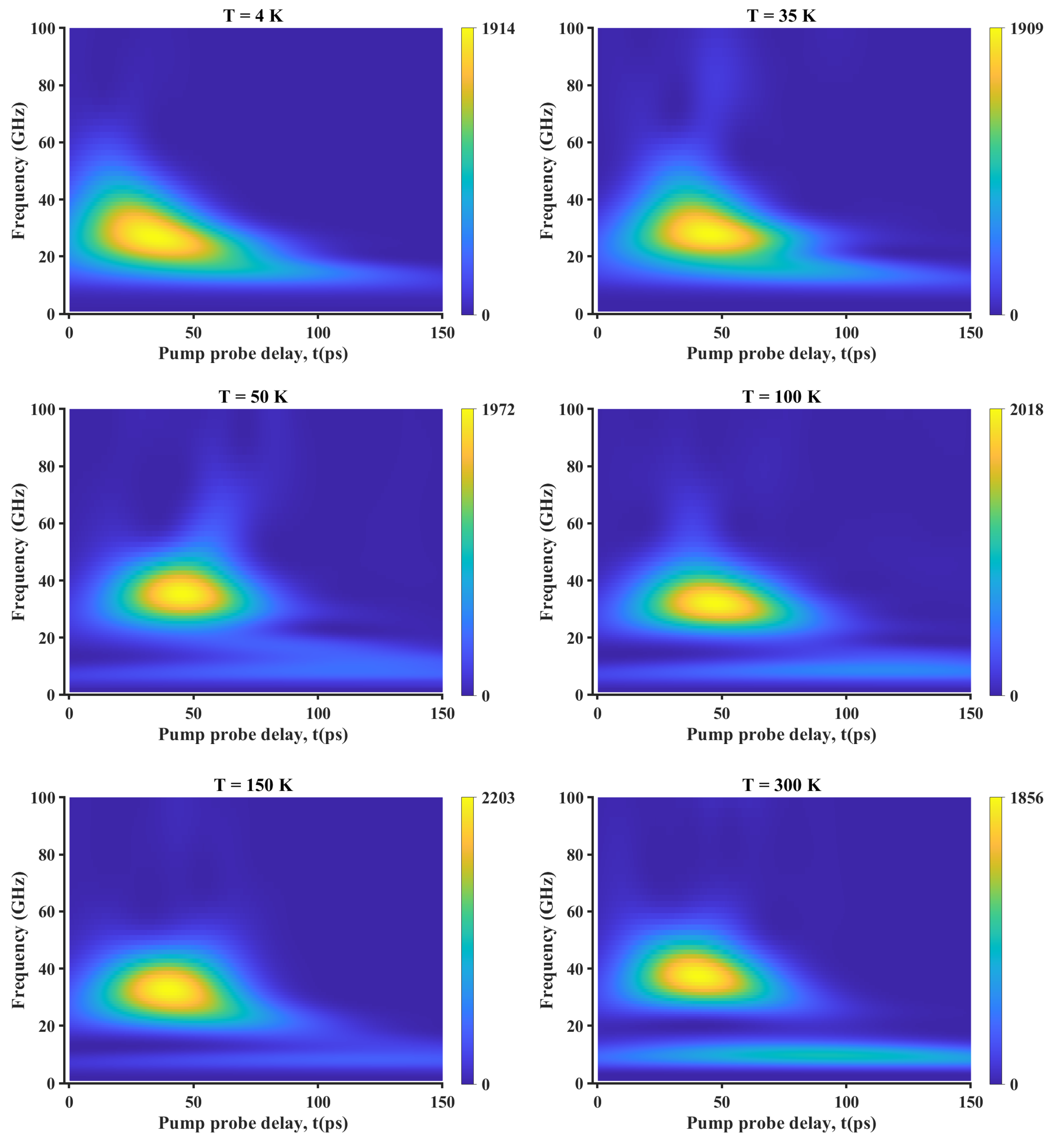}
\caption{CWT spectra at various temperatures: The evolution of the CWT amplitude  with temperature at different pump-probe delay times. For the clarity, CWT spectra at the temperatures 4 K, 35 K, 50 K, 100 K, 150 K and 300 K are shown here.}\label{T_dependent_CWT}
\end{figure}

Figure \ref{T_dependent_CWT} illustrates the evolution of the CWT spectrum (frequency spectrum of the strain pulse) as a function of  pump probe delay  at various temperatures obtained from continuous wavelet transform.

\newpage

\section{ Phenomenological theoretical model}

A lattice model of phonons is described by the Lagrangian,

\begin{equation}
\mathcal{L} = \sum_{j} \left[ \dot{Q_{j}}^{2} - \lambda(Q_{j+1} - Q_{j})^{2}  \right] \label{L1} 
\end{equation}

where $Q_{j}$ is the distortion of the $j^{th}$ atom away from its equilibrium position and $\lambda$ is the spring constant of the restoring potential. The distortions $Q_{j}(t)$ are functions of the position index j as well as time t. At this point, we introduce three additional terms in the Lagrangian that represent coupling between the lattice distortions and the magnetic interactions. The magnetic interactions are taken to be of the nearest-neighbor kind: $\vec{S}_{j} \cdot \vec{S}_{j+1}$.  On phenomenological grounds, we allow the magnetic interactions to couple to the linear, quadratic, and gradient components of the distortion, represented by $Q_{j}$ , $Q_{j}^{2}$ and $(Q_{j+1} -  Q_{j})^{2}$, through the coupling constants $\gamma$,  $\xi$ and $\chi$, respectively.

\begin{equation}
\mathcal{L} = \sum_{j} \left[ \dot{Q_{j}}^{2} - \lambda(Q_{j+1} - Q_{j})^{2} + \gamma \vec{S}_{j} \cdot \vec{S}_{j+1} Q_{j}  - \xi \vec{S}_{j} \cdot \vec{S}_{j+1} Q_{j}^{2} + \chi \vec{S}_{j} \cdot \vec{S}_{j+1}(Q_{j+1} - Q_{j})^{2}  \right] \label{L2}
\end{equation}

We treat the magnetic interactions at a mean - field level, replacing the magnetic interactions with their expectation value.  From the onset of the LRMO, the expectation value  \textlangle $\vec{S}_{j} \cdot \vec{S}_{j+1}$\textrangle acquires a uniform non-zero value M$^{2}$(T), T being the temperature. Above the T$_{3D}$, M$^{2}$(T) is zero. For the LRMO phase, the Lagrangian takes the form

\begin{equation}
\mathcal{L} = \sum_{j} \left[ \dot{Q_{j}}^{2} - (\lambda - M^{2}\chi) (Q_{j+1} - Q_{j})^{2} + M^{2}\gamma Q_{j} - M^{2}\xi Q_{j}^{2} \right] \label{L3}
\end{equation}

\begin{equation}
 = \sum_{j} \left[ \dot{Q_{j}}^{2} - (\lambda - M^{2}\chi) (Q_{j+1} - Q_{j})^{2} - M^{2}\xi \left(Q_{j} - \frac{\gamma}{2\xi}\right)^{2} \right] \label{L4}
\end{equation}
where we have neglected the overall constant. 
We define a modified distortion $\bar{Q_{j}}$ = $Q_{j} - \frac{\gamma}{2\xi}$. In terms of the modified distortion the Lagrangian becomes, 

\begin{equation}
\mathcal{L} = \sum_{j} \left[ \bar{\dot{Q_{j}}}^{2} - (\lambda - M^{2}\chi) (\bar{Q}_{j+1} - \bar{Q}_{j})^{2} - M^{2}\xi \bar{Q}_{j}^{2} \right] \label{L5}
\end{equation}

At this point, one can take the continuum limit of the distortions $\bar{Q}$ being much smaller than the lattice spacing, transforming the discrete index j to the continuum variable x. The distortion $\bar{Q}$(x,t) then becomes a classical field $\bar{Q}$(x, t), and the Lagrangian becomes

\begin{equation}
\mathcal{L} = \int_{0}^{L} dx\left[ (\partial_{t}{\bar{Q}(x,t))}^{2} - (\lambda - M^{2}\chi)(\partial_{x}{\bar{Q}(x,t))}^{2} - M^{2}\xi \bar{Q}(x,t)^{2} \right] \label{L6}
\end{equation}

The equation of motion for this Lagrangian is 

\begin{equation}
\Ddot{\bar{Q}} = \left[ (\lambda - M^{2}\chi) \partial_{x}^{2} - M^{2}\xi \right]\bar{Q} \label{L7}
\end{equation}

We now introduce the Fourier decomposition of the field $\bar{Q}$(x,t) in momentum (k) and frequency ($\omega$) domain in terms of the dual models $\Tilde{Q}(k,\omega)$:

\begin{equation}
\bar{Q}(x,t) = \int dk ~ d\omega ~ e^{-ikx -i\omega t} \Tilde{Q}(k,\omega) \label{L8}
\end{equation}

In terms of the dual models, the equation of motion becomes

\begin{equation}
\left[ -\omega^{2} + (\lambda - M^{2} \chi) k^{2} + M^{2} \xi\right] \Tilde{Q}(k,\omega) = 0 \label{L9}
\end{equation}

This equation also reveals the dispersion of the phonons

\begin{equation}
\omega(k) = \sqrt{(\lambda - M^{2}\chi)k^{2} + M^{2} \xi} \label{main_dispersion}
\end{equation}

where k is the phonon wave-vector. As a consequence, when M is zero (for T $>$ T$_{3D}$), the dispersion simplifies to the expected linear form for acoustic phonons: $\omega(k)$ = $\sqrt{\lambda} k$. When M and the coupling $\chi$ are nonzero (for T $<$  T$_{3D}$) but the coupling $\xi$ is negligible, the dispersion retains its linear form. However, the effective spring constant $\lambda$ is reduced, leading to a softening of the phonon mode.

\begin{equation}
\omega(k) = \sqrt{\lambda_{eff}} k, ~ \lambda_{eff} = \lambda - M^{2}\chi .
\end{equation}

For the case M$^2\chi < \lambda$, the coupling between the spin-spin correlations and the gradient component of the distortions leads to the softening of the HFAP across T$_{3D}$ . 

The interplay between spin-spin correlations and lattice distortions plays a crucial role in the gapping out the LFAP below a  temperature, T$_{3D}$. This is captured by the dispersion relation (equation (\ref{main_dispersion})) when all interaction parameters (M, $\chi$ and $\xi$) are non-zero.

\section{~~Estimation of coupling constant values}	
\subsection{ Estimation of $\chi$ from HFAP frequency shift}	

Between 50 K and 35 K, the central frequency of the HFAP exhibits a 6 GHz decrease in frequency, dropping from 33 GHz to 27 GHz (as shown in Figures 3(b) and 3(c) of the main manuscript). This shift in frequency provides a means to estimate the value of $\chi$, the coupling constant between the spins and the gradient components of the lattice distortion.

For temperature T $>$ T$_{3D}$ = 50 K, the dispersion relation for the acoustic phonons in the HFAP takes the form given by, 

 \begin{equation}
\omega_{50}(k) = \sqrt{\lambda} k \label{L11}
\end{equation}
here $\omega_{50}(k)$, represents the frequency at T = 50 K.  For temperature T $<$ T$_{3D}$ = 50 K, the dispersion relation takes the form given by, 

\begin{equation}
\omega_{35}(k) = \sqrt{(\lambda - M^{2}\chi)k^{2}} \label{HFAP_dispersion_function}
\end{equation}

for negligible $\xi$, the coupling constant between the spins and the quadratic components of the lattice distortion. Here $\omega_{35}(k)$, represents the frequency at T = 35 K.  Taking the ratio between the above equations(18 and 19), we get,
 \begin{equation}
\frac{\omega_{50}}{\omega_{35}} = \frac{\sqrt{\lambda}}{\sqrt{(\lambda - M^{2} \chi)}}
\end{equation}

Substituting $\frac{\omega_{50}}{\omega_{35}}$  = $\frac{33 GHz}{27 GHz}$, $\chi$ is found to be, 

 \begin{equation}
\chi \approx  \frac{\lambda}{3 M^{2}}
\end{equation}

With a density of 5.24 g/cm$^{3}$ and  a bulk modulus of 22 GPa \cite{jain2013commentary}, the estimated speed of sound in  CrSiTe$_{3}$ is $v_{HFAP}$ = $ \sqrt{\lambda}$ $\sim$ 2.1 $\times$ $10^{3}$ m/s. Using the measured magnetization value at T = 35 K, i.e., M = 28.82  $\times$ $10^{3}$ A/m, the value of $\chi$ is found to be $\sim$ 1.7 mHz $m^{2}/ A^{2}s$.

Figure \ref{HFAP_dispersion} compares the frequency dispersion  of HFAP at two distinct temperatures: 50 K and 35 K. In the figure, the red-shift of $\sim$ 6 GHz is shown at k$\sim 9.9 \times 10^5$ cm$^{-1}$ for parameters of $\chi$  $\sim$ 1.7 mHz $m^{2}/ A^{2}s$ and negligible $\xi$,  highlighting the significant influence of coupling constant $\chi$ on the sudden softening of the central frequency of HFAP.

\begin{figure}%
\includegraphics[width= 1\textwidth]{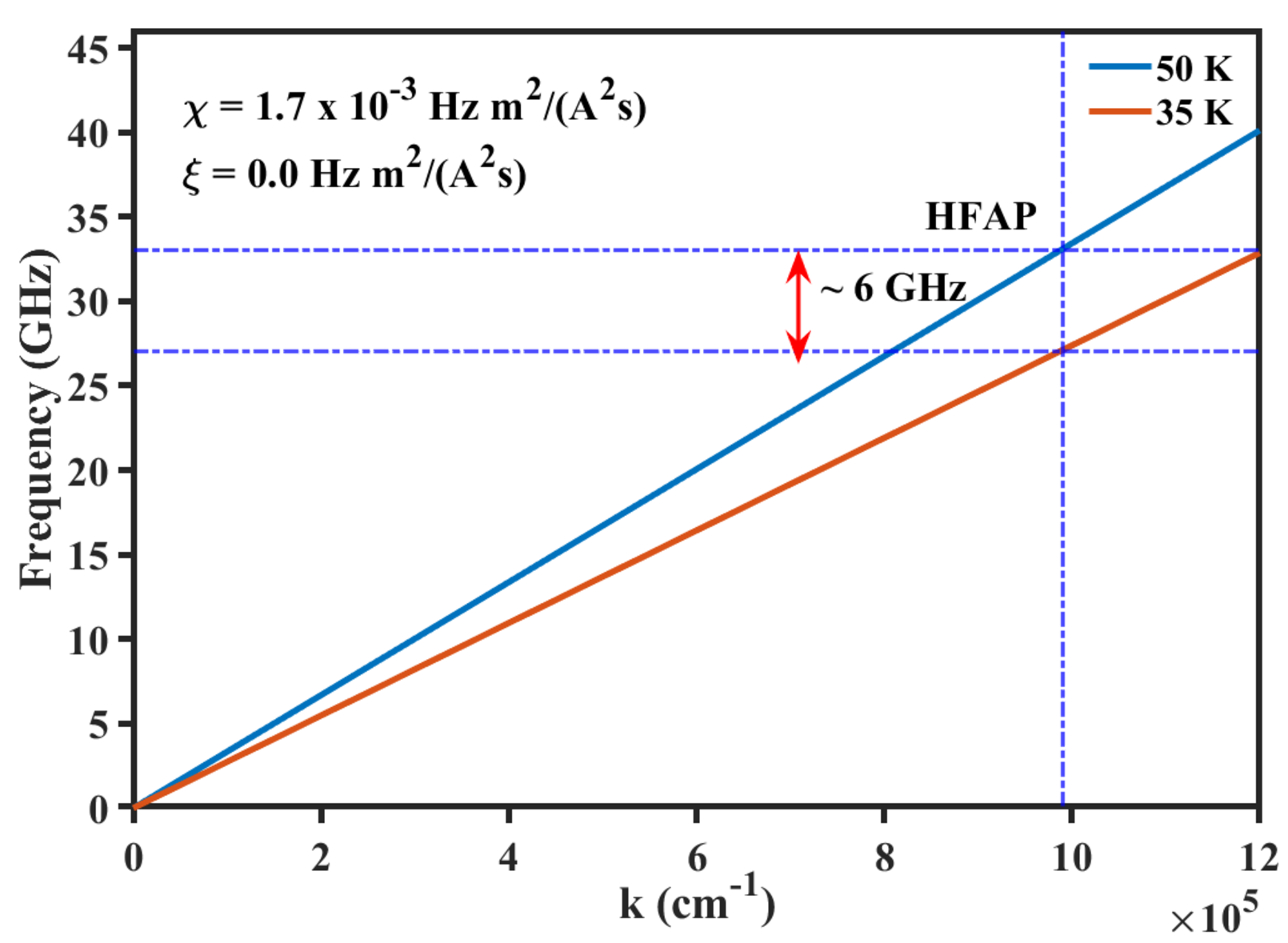}
\caption{ Frequency  dispersion of HFAP: This figure shows the dispersion relation of HFAP at two different temperatures: 50 K (blue solid line) and 35 K (red solid line). The shift in the frequency of HFAP  is marked by the dark blue dashed lines and indicated by the red double arrow . The parameters $\chi$ and $\xi$ used to generate the dispersion spectrum at T = 35 K using equation (\ref{HFAP_dispersion_function}) are also shown on the top left corner of the plot.}\label{HFAP_dispersion}
\end{figure}

\newpage

\subsection{ Estimation of $\xi$ from LFAP frequency shift}
The central frequency of the LFAP at 35K exhibits a blue shift of 7 GHz from its value at 50 K (from 9.6 GHz to 16.6 GHz; Figures 3(b,c)), providing a key tool for estimating $\xi$.

Since the central frequency of the LFAP (9.6 GHz) at T = 50 K is lesser than the central frequency of the HFAP (33 GHz), the speed of the sound is scaled accordingly. 
\begin{equation}
    v_{LFAP} = 0.29*v_{HFAP} \label{speed}
\end{equation}

The dispersion relation for the acoustic phonons in LFAP below T $<$ T$_{3D}$ = 50 K, takes of the form,  

\begin{equation}
\omega(k) = \sqrt{(\lambda - M^{2}\chi)k^{2} + M^{2} \xi}
\end{equation}
and using the similar approach mentioned above and for the negligible value of $\chi$, the value of the  $\xi$ is estimated using 

\begin{equation}
\xi \approx \frac{0.66 \lambda k^{2}}{M^{2}} 
\end{equation}

Using suitable values, $\xi$ is found to be $\sim$ 10.2 THz $m^{2}/ A^{2}s$. 

Figure \ref{HFAP_dispersion} compares the dispersion relation of the acoustic phonons in LFAP at two distinct temperatures: 50 K and 35 K. A blue shift of $\sim$ 7 GHz in the LFAP frequency is shown by vertical double headed arrow. This shift is accurately reproduced using parameters of $\chi$  $\sim$ 10.2 THz $m^{2}/ A^{2}s$ and negligible $\xi$ (here $\chi$  $\sim$ 1.7 mHz $m^{2}/ A^{2}s$ has  a negligible effect on the $\omega(k)$ ). This highlights the  significant influence of coupling constant $\xi$ on the  blue shift of the LFAP .

\begin{figure}%
\includegraphics[width= 1\textwidth]{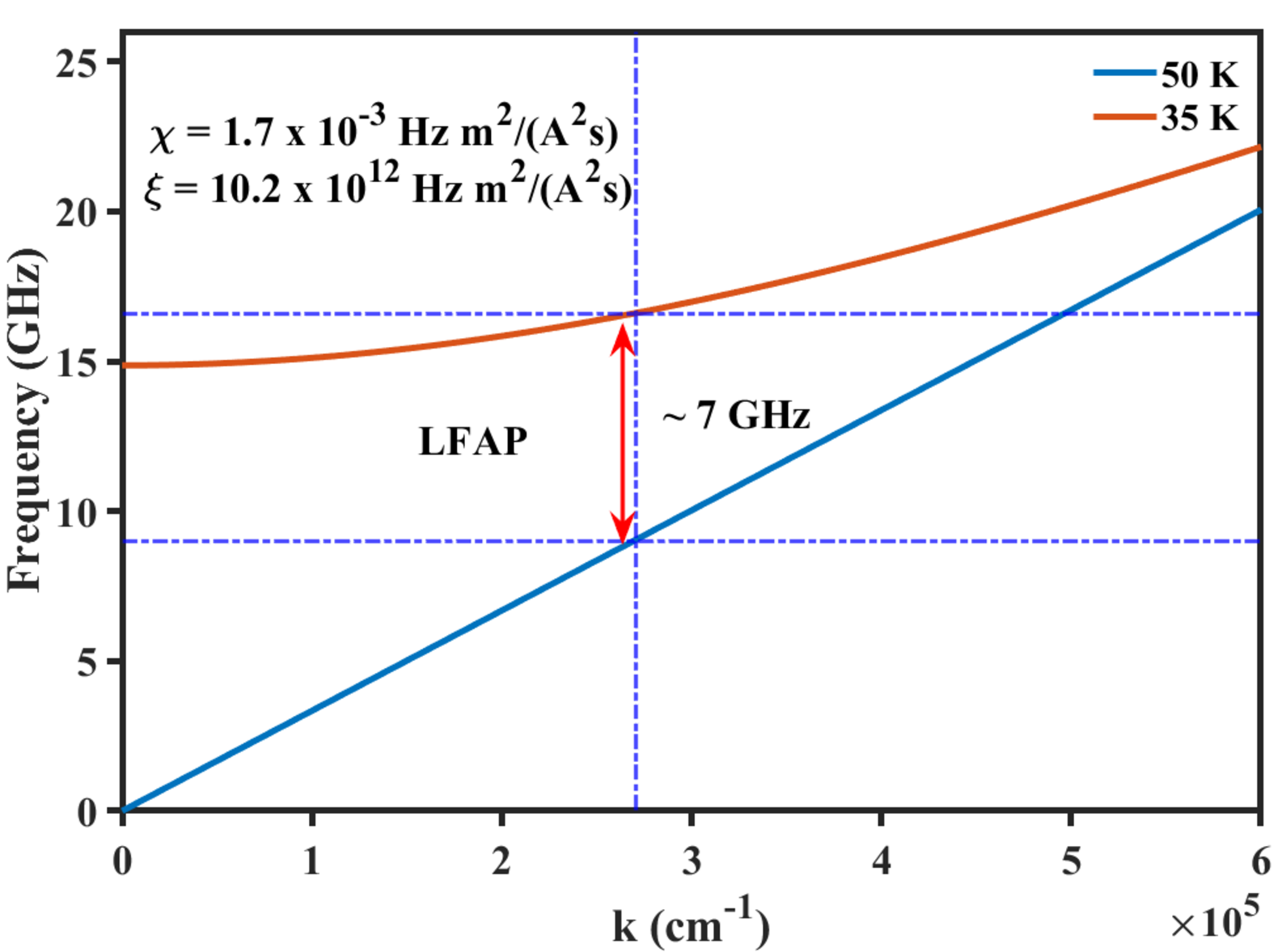}
\caption{ Phonon dispersion relation of the LFAP: This figure shows the dispersion relation of acoustic phonons of LFAP at two different temperatures: 50 K (blue solid line) and 35 K (red solid line). The  shift in the frequency of LFAP  is marked by the dark blue dashed lines and indicated by the red double arrow. The parameters $\chi$ and $\xi$ used to generate the dispersion spectrum at T = 35 K are also shown at the top left corner.}\label{HFAP_dispersion}
\end{figure}

Observe that the central frequency of the high-frequency acoustic component exhibits anharmonic decay behavior within the temperature range of T=50 K - 300K, and a notable softening across T$_{3D}$. Concurrently, the low-frequency acoustic component demonstrates minimal softening within the 50K-300K range, but exhibits a contrasting blue shift across T$_{3D}$. This implies that both acoustic components could originate from two distinct branches of acoustic phonon. Hence the speed of sound can be different for HFAP and LFAP as in equation \ref{speed}. 

\newpage

\section{~~ Data fitting}	

The commercial software MATLAB provides many numerical algorithms for solving unconstrained minimization problems. We used the simplex algorithm based on the Nelder-Mead simplex method, which is implemented in MATLAB as the function fminsearch.

We fit the experimental data DR to a multi-exponential function, DR$_{multi}$, to extract the desired parameters. We have used either a bi- or tri-exponential function, as defined in equations \ref{tri_exp} and \ref{Bi_exp}, respectively. We minimized the error function S = DR - DR$_{multi}$ to find the best fit. The data and its fit are shown in Fig. \ref{Multi_exp_fit} (a) and (b).
To calculate the error bars, we have used orthogonal distance regression (ODR) approach \cite{boggs1992user, watson2005odrpack95}. The orthogonal ODR algorithm minimizes error function by adjusting both fitting parameters and values of the independent variable in the iterative process. The error bars on the fit parameters are spit out automatically in MATLAB code, we implemented this by calculating the Jacobian matrix followed by the Variance-Covariance Matrix for the fit parameters.

\subsection{~~ Goodness of the fit:} 
The quality or the goodness of the fit was assessed using both the graphical method and  the coefficient of the determination Chi-squared (R$^{2}$), which is given by 

\begin{equation}
R^{2} = \frac{\sum(DR - DR_{multi})^{2}}{\sum(DR - \textlangle DR \textrangle)^{2}}
\end{equation}

\end{document}